\documentclass[aps,floats,twocolumn,prb,showpacs]{revtex4-1}

\usepackage{amsmath}
\usepackage{graphicx}
\usepackage{xcolor}
\usepackage{epstopdf}
\usepackage{natbib}
\usepackage{xfrac}
\usepackage[normalem]{ulem}
\usepackage{multirow}
\usepackage{physics}
\usepackage{amssymb}

\usepackage{verbatim}

\newcommand{\mw}[1] {\textcolor{black}{#1}}
\newcommand{\mb}[1] {\textcolor{black}{#1}}
\newcommand{\new}[1] {\textcolor{black}{#1}}
\newcommand{\old}[1] {}
\newcommand{\me}[1] {\textcolor{black}{#1}}

\begin{document}
\title{\new{Quantum Boltzmann equation for strongly correlated systems: comparison to dynamical mean field theory}}



\author{M.~Wais$^1$}
\email[]{michael.wais@tuwien.ac.at}
\author{M. Eckstein$^2$}
\author{R. Fischer$^1$}
\author{P. Werner$^3$}
\author{M. Battiato$^4$}
\author{K. Held$^1$}
\affiliation{$^1$ Institute of Solid State Physics, TU Wien,  Vienna, Austria}
\affiliation{$^2$ Department of Physics, University of Erlangen-N\"urnberg, 91058 Erlangen, Germany}
\affiliation{$^3$ Department of Physics, University of Fribourg,  Fribourg, Switzerland}
\affiliation{$^4$ Nanyang Technological University, 21 Nanyang Link, Singapore, Singapore}

\begin{abstract}

We investigate the potential of a \new{quantum} Boltzmann equation without momentum conservation 
for description of strongly correlated electron systems out of
equilibrium. 
\new{ In a spirit similar to dynamical mean field theory (DMFT),
the momentum conservation of the electron-electron
scattering is neglected, which yields a time-dependent occupation function
for the equilibrium spectral
function, even in cases where
well-defined quasiparticles do not exist. The main assumption of this method is that the spectral function remains sufficiently rigid under the non-equilibrium evolution.} 
 We compare the result of the \new{quantum} Boltzmann equation to
non-equilibrium DMFT simulations for the case of photo-carrier relaxation in Mott insulators, where processes
on very different timescales emerge, i.e., impact ionization, intra-Hubbard-band thermalization, and
full thermalization.  Since \new{quantum} Boltzmann simulations without momentum conservation are computationally cheaper than non-equilibrium DMFT, this method allows the simulation of more complicated systems or devices, and to access much longer times.

\end{abstract}

\date{\today}

\maketitle
\section{Introduction}

The description of  excited quantum many-body systems is among the most difficult tasks in computational physics. Numerous methods have been developed that employ approximations on different levels in order to make the solution of the many-body problem feasible. 
Starting from the Keldysh formalism,\cite{Keldysh1964} the most prominent approaches are based on either quantum kinetic equations or non-equilibrium Green's function (NEGF) techniques.\cite{Rammer1986,haug1998quantum}
Quantum kinetic equations, which generalize the classical Boltzmann equation,\cite{Boltzmann1872} have a long and successful history in the description of the dynamics of semiconductors.\cite{haug1998quantum} In the most straightforward derivation of a Boltzmann equation, one usually assumes the existence of  quasiparticles with a well-defined dispersion $\epsilon_{\bold k}$; the Boltzmann equation then describes the evolution of the quasiparticle occupations, while the quasiparticle bands change at most in a mean-field sense. Quantum effects are contained via the Fermi-Dirac statistics of the particles, but the phase information between individual scatterings is lost.

In strongly correlated systems, the assumption of well-defined quasiparticles is not justified in many situations. In particular, doped Mott insulators in equilibrium display bad metallic behavior above some rather low coherence temperature, with a scattering length of the order of the lattice spacing,\cite{Deng2013} and also photo-doped Mott insulators do not quickly relax to a Fermi liquid.\cite{Eckstein2013,Sayyad2016} Furthermore, in correlated systems, the electronic spectrum itself can depend strongly on the non-equilibrium distribution, 
which may result in 
a photo-induced renormalization \cite{Wegkamp2013,Golez2015,Golez2017} or filling of the Mott gap, 
similarly to what happens by increasing temperature.\cite{Mo2004}
For these reasons, the dynamics of correlated systems is often 
studied using 
formally exact 
but computationally expensive 
non-equilibrium Green's function (NEGF) techniques. In this approach, the self-energy acts as a memory kernel for the propagation of the Green's function $G_{\bold k}(t,t')$, which contains information about both the spectrum and the occupation. Depending on the physics to be studied, NEGF techniques are combined with different diagrammatic approximations, including weak-coupling approximations such as the GW or second Born approximation \cite{Sentef2013,Golez2016,Schluenzen2017} and the functional renormalization group,\cite{Jakobs2007,Kennes2012} or,  in the strongly correlated regime, dynamical mean field theory (DMFT).\cite{Aoki2014,Georges1996}

In the latter case, a particular problem is the lack of efficient and reliable impurity solvers.  
Non-perturbative techniques such as  exact diagonalization,\cite{alvfeh11,alvermanetal12,Gramsch2013} matrix-product state methods,\cite{Wolf2014,Balzer2015,Ganahl2015} or continuous-time quantum Monte Carlo\cite{Werner2009b} face severe limitations when applied to real-time dynamics and typically work only for short times or, alternatively, in steady states.\cite{Arrigoni2013,Dorda2014} Even approximate solvers such as the non-crossing approximation \cite{Eckstein2010} are restricted to short time scales. 
The restriction of NEGF techniques to short times makes it difficult to describe situations where relevant processes take place on very different timescales. A prime example are relaxation processes in photo-excited Mott insulators, which include ultra-fast spin-charge relaxation processes,\cite{Eckstein2016SciRep,Golez2014} impact ionization,\cite{Werner2014,Sorantin2018} and recombination,\cite{Eckstein2011} ranging from few to thousands of hopping times.

It would therefore be desirable to have a generalization of a kinetic equation which can describe the dynamics in correlated systems such as Mott insulators, in spite of the fact that 
well-defined quasiparticles may not exist. 
In fact, on the qualitative level one often argues in terms similar to a Boltzmann equation when explaining the ``transfer of occupations'' between different energy windows.\cite{Werner2014} 
Here we attempt to put such arguments on a more solid basis. 
\mb{We derive a quantum Boltzmann equation from the interacting Green's function in the DMFT limit, and show that it has the same mathematical structure as the quasiparticle Boltzmann equation in the limit of infinite dimensions (i.e. negligible momentum- ($\bold k$-) conservation). We then solve numerically the derived quantum Boltzmann equation and compare it to non-equilibrium DMFT results, showing that the two methods yield qualitatively and semi-quantitatively similar results, even if we use a rather crude approximation of the scattering amplitude within the quantum Boltzmann equation.} 

\new {The paper is organized as follows:} In Section \ref{chap:Boltz}, we \new{derive} the Boltzmann equation \mb{in the limit of infinite dimensions. To ease a later comparison and help the reader's intuition, we first show, in Section \ref{secqbewdhmv}, the structure of the quasiparticle Boltzmann equation (QPBE) in the limits of infinite dimensions, by neglecting the momentum-dependence in the collision term. Here, the basic assumption is that, even if electronic correaltions strongly modify the spectral function, the basic excitations are still quasiparticle-like and their dynamics subject to the Boltzmann equation.}
In Section \ref{secqbe} we derive a quantum Boltzmann equation (QBE), starting with the non-equilibrium DMFT equations. This more rigorous proof does not require the existence of well defined quasiparticles but only the rigidity of the spectral function. 
The 
final structure of the two Boltzmann equations is the same, but the QBE  holds under more general circumstances. In  Section \ref{chap:laser} we develop two different approaches on how to account for the laser excitation: (i) Fermi's golden rule and (ii) including the effects of a finite number of laser oscillations and amplitude modulations in first order perturbation theory. In  Section \ref{chap:dos} we report how we parametrise the quantum Boltzmann equation. Numerical details are sketched in Section \ref{chap:numImp}.

%
%
The results of our numerical simulations are presented in Section \ref{chap:results}. We \new{focus on}\old{describe} the
relaxation dynamics of doublons in a photo-excited Mott insulator in Section \ref{chap:twotimescales}, and
discuss the 
different ways to model the laser excitation
in Section \ref{chap:laserEx}. The population
dynamics in the upper and lower Hubbard band is presented in  Section
\ref{chap:popDyn} and compared to non-equilibrium DMFT.
 Section \ref{chap:scatStr} 
briefly discusses the scattering amplitude,
the only adjustable parameter of the theory, while Section
\ref{chap:threeStep} reveals the three steps in the thermalization 
process: (i) impact ionization, (ii) thermalization within the two
Hubbard bands, (iii) full thermalization between the Hubbard bands.
Finally,  Section \ref{chap:concl}  summarizes our results \new{and provides an outlook}.

\section{(Quantum) Boltzmann equation in the limit of $\infty$ dimensions} \label{chap:Boltz}



In this section, we derive a Boltzmann equation which can qualitatively capture the non-equilibrium DMFT dynamics of strongly correlated systems in certain limits.  In contrast to previous work,\cite{Stark2013} we describe the time-dependence of the distribution function of the {\em interacting} equilibrium spectrum via the Boltzmann equation---not the non-interacting one. The essential approximation is that this spectrum remains rigid under  non-equilibrium excitations.  
\mb{Before addressing the quantum Boltzmann equation (QBE) in the DMFT limit, we analyze the quasiparticle Boltzmann equation (QPBE) in the same limit.
}
  
The QPBE \new{assumes the existence of quasiparticles} and describes the time-evolution of the quasiparticle distribution function $f(\bold r, \bold k , t)$, i.e., the probability that a quasi-electron wave-packet state centered at phase-space point $(\bold r, \bold k )$ is occupied at time $t$. It is given by the Fermi distribution $f(\bold k)=1/(e^{\beta(\epsilon_{\bold k}-\mu)}+1)$ in equilibrium, with the single-particle dispersion relation $\epsilon_\bold{k}$, the inverse temperature $\beta$ and the chemical potential $\mu$. The QPBE can capture many aspects of the DMFT dynamics in the limit of weak interactions,\cite{Stark2013} but its applicability seems a priori unclear for strongly correlated systems. \new{Here, the poles of the one-particle Green's function still describe one-particle excitations which might be however
broadened considerably by a finite imaginary part of the self-energy. Considering a quasi-particle distribution function in energy (frequency) 
$f({\mathbf r},\omega,t)$ certainly only makes sense if the line-width given by the \new{imaginary part of the self-energy} is considerably smaller than the bandwidth.}
 
On the other hand, \mb{we will show that} one can write a quantum Boltzmann equation for a distribution function {$F(\bold r, \bold k , t, \omega)$,} which is related to the ratio of the occupied density of states to the spectral function, \mb{ and can be simplified to  $F(\bold r,  t, \omega)$ in the DMFT limit}. The latter can be defined even in the absence of well-defined quasiparticles, and gives $F(\omega)=1/(e^{\beta (\omega - \mu)}+1)$ in equilibrium (a precise definition will follow below). We will refer to this approach as the quantum Boltzmann equation (QBE), although in the literature the term QBE is sometimes also used for the QPBE.

\mb{ 
Both equations turn out to have the same mathematical structure, which is a local Boltzmann equation, although the meaning of the distribution function is different in the two cases\new{, i.e., more general for the QBE}. Hence this local Boltzmann equation can potentially capture many aspects of the DMFT evolution not only in the limit of weak interactions, but also for strongly correlated Mott insulators where the concept of  well-defined quasiparticles is questionable. }

\subsection{Quasiparticle Boltzmann equation in the DMFT limit}
\label{secqbewdhmv}

The Boltzmann equation for the time-evolution of the quasiparticle distribution function $f(\bold r, \bold k , t)$ reads \cite{Ziman1960}
\begin{equation}
\frac{\partial f}{\partial t} + \bold v_r \cdot \nabla_{\bold r} f + \bold v_k \cdot \nabla_{\bold k} f = \left (\frac{\partial f}{\partial t } \right ) _\textrm{col} \textrm{ .} \label{eq:boltzmann}
\end{equation}
The quantities $\bold v_r$ and $\bold v_k$ are the velocities of a single-particle wave-packet in real- and momentum-space and can be derived from the single-particle Schr{\"o}dinger equation as \footnote{Here, the spread of the wave packet is disregarded and the phase information is not considered.}
\begin{subequations}
\begin{align}
\bold v_r (\bold k)  &  = \frac{1}{\hbar} \nabla_{\bold k} \epsilon(\bold k)+ \bold v_{\text{an}} (\bold k) \textrm{ ,}\\
\bold v_k (\bold r, \bold k,t) &  =  \frac{-e}{\hbar} \left ( \bold E(\bold r , t) + \frac{1}{c} \bold v_r (\bold k) \times \bold B(\bold r , t)  \right ) \textrm{ ,} 
\end{align}
\end{subequations}
with the absolute value of the electron charge $e$, the reduced Planck-constant $\hbar$, the speed of light $c$, the single-particle dispersion relation $\epsilon(\bold k)$, the macroscopic electric and magnetic fields $\bold E (\bold r , t)$, $\bold B (\bold r , t)$, and the anomalous component of the wavepacket group velocity $\bold v_{\text{an}} (\bold k)$. \cite{DiXiao2010} Since we only discuss systems with zero $\bold E$- and $\bold B$-fields in this paper, the third term on the left-hand side of Eq.~\eqref{eq:boltzmann} vanishes. Furthermore, we only allow spatially uniform electron populations, i.e. $\nabla_{\bold r} f=0$, which implies that the second term vanishes as well, and only the first term $\frac{\partial f}{\partial t}$ survives. 
For better readability we will suppress the explicit time dependence of $ f$ from now on unless it is needed for a better understanding. 
The term on the right-hand side of Eq.~\eqref{eq:boltzmann} is the so-called collision term that describes 
all the scattering processes of the electrons, i.e. electron-electron, electron-phonon and electron-impurity scattering. Here we focus on electron-electron scattering for which the scattering term reads\cite{Ziman1960}
\begin{equation}
\begin{split}
& \left (\frac{\partial f (\bold k_0 )}{\partial t } \right ) _\textrm{col-ee} = \sum_{\bold G} \int \mathrm d ^d k_1 \mathrm d ^d k_2 \mathrm d ^d k_3  \Big [ \\
& \quad - W_{\bold G}(\bold k_0,\bold k_1,\bold k_2,\bold k_3) f(\bold k_0) f(\bold k_1) (1-f(\bold k_2)) (1-f(\bold k_3))\\
 &\quad + W_{\bold G}(\bold k_2,\bold k_3,\bold k_0,\bold k_1)  (1-f(\bold k_0)) (1-f(\bold k_1)) f(\bold k_2) f(\bold k_3) \Big ],
\end{split} \label{eq:col}
\end{equation}
where $d$ is the spatial dimension of the system and the scattering amplitude is
\begin{equation}
\begin{split}
W_{\bold G} (\bold k_0,\bold k_1,\bold k_2,\bold k_3) = &w(\bold k_0,\bold k_1,\bold k_2,\bold k_3) \\
& \times \delta \left ( \bold k_0 + \bold k_1 - \bold k_2 - \bold k_3  + \bold G \right ) \\
& \times \delta \left ( \epsilon(\bold k_0) + \epsilon(\bold k_1) - \epsilon(\bold k_2) - \epsilon(\bold k_3)  \right ) \textrm{ .}
\end{split}
\label{eq:scatAmpl}
\end{equation}
The vector $\bold G$ is a reciprocal lattice vector and $w(\bold k_0,\bold k_1,\bold k_2,\bold k_3)$ is the scattering probability that gives the probability for a process where two particles with momenta $\bold k_0$ and $\bold k_1$ scatter and end up with $\bold k_2$ and $\bold k_3$ (or for the inverse process). 
The two $\delta$-functions ensure momentum- and energy-conservation. 

It is non-trivial to obtain $w(\bold k_0,\bold k_1,\bold k_2,\bold k_3)$ ab-initio for a real material as it corresponds to the effective interaction. We assume that $\epsilon(\bold k)$ is already the renormalized dispersion relation and only the residual (i.e. screened) interaction enters into the scattering probability. Furthermore we assume an effective scattering probability (or effective interaction) that does not depend on the momenta, i.e.\mw{
\begin{equation}
w(\bold k_0,\bold k_1,\bold k_2,\bold k_3) \to \frac{\alpha}{{V_{BZ}}^2} = \textrm{const.}
\end{equation}
with the volume of the Brillouin zone $V_{BZ}$ and a constant $\alpha$.}
Another difficulty in calculating Eq.~\eqref{eq:col} is the fact that it includes two $\delta$-functions. While the momentum-$\delta$ could be resolved relatively easily, as it simply fixes one of the three integration variables $\bold k_i$\new{,} the energy-$\delta$ is more difficult, since $\epsilon (\bold k)$ is an arbitrary function. DMFT methods become exact in the limit of infinite dimensions,\cite{Metzner1989} where the influence of $\bold k$-conservation vanishes. When DMFT is used as an approximation, the $\bold k$-conservation is also dropped when the method is applied to a finite-dimensional system. In analogy to DMFT we therefore give up the momentum conservation of the Boltzmann equation by assuming\mw{
\begin{equation}
\sum_\bold G \delta (\bold k_0 +\bold k_1 - \bold k_2 - \bold k_3 + \bold G)   \to \frac{1}{V_{BZ}} = \textrm{const.},
\end{equation}}
where the factor $\frac{1}{V_{BZ}}$  restores the correct unit and order of magnitude.
After we have given up momentum conservation, we also do not have to keep the momentum resolution of the distribution function. Instead we consider the probability that a state at a certain energy (rather than a certain momentum) is occupied, i.e. $f(\bold k) \to f(\epsilon)$. With all the above simplifications Eq.~\eqref{eq:boltzmann} reads
\begin{align}
\frac{\partial f (\epsilon_0 )}{\partial t }  =
&\frac{\alpha}{{V_{BZ}}^3}  \int  \mathrm d ^d k_1 \mathrm d ^d k_2 \mathrm d ^d k_3 \, \delta \left ( \epsilon_0 + \epsilon_1 - \epsilon_2 - \epsilon_3  \right )  
\nonumber\\
&  \times 
\mathcal{P}[f(\epsilon_0),f(\epsilon_1),f(\epsilon_2),f(\epsilon_3)],
\label{eq:col2}
\end{align}
where $\epsilon_j\equiv \epsilon (\bold k_j)$, and we have introduced the phase-space factor
\begin{align}
\label{gdjadsljbja}
\mathcal{P}&[f_0,f_1,f_2,f_3]
\nonumber\\&=
(1-f_0)(1-f_1)f_2f_3-(1-f_2)(1-f_3)f_0f_1,
\end{align}
\new{which yields   $\frac{\partial f (\epsilon_0 )}{\partial t }  = 0$ if $f_i$ is the (equilibrium) Fermi function.}

At this point the integration is still performed over the momenta,  while the integrand only depends on the momenta through the dispersion relation. We can therefore reduce the integrals into an energy integration by introducing the density of states $\rho(\epsilon)=\frac{1}{V_{BZ}} \int  \mathrm d ^d k\,\delta(\epsilon-\epsilon(\bold k))$, so that
the collision term Eq.~\eqref{eq:col2} becomes (cf.~Ref.~\onlinecite{Stark2013})
\begin{align}
&\frac{\partial f (\epsilon_0 )}{\partial t }  =
\alpha\! \!\int   \mathrm d  \epsilon_1 \mathrm d  \epsilon_2 \mathrm d  \epsilon_3 \delta \left ( \epsilon_0\! +\! \epsilon_1\! -\! \epsilon_2 \!-\! \epsilon_3  \right ) 
\,\,\nonumber\\
& \,\,\,\,\,\times 
\rho(\epsilon_1) \rho(\epsilon_2) \rho(\epsilon_3) 
 \mathcal{P}[f(\epsilon_0),f(\epsilon_1),f(\epsilon_2),f(\epsilon_3)].
  \label{eq:col3} 
\end{align}
\mw{Equation \eqref{eq:col3} defines the local Boltzmann equation, 
which is much easier to solve numerically than Eq.~\eqref{eq:col} since the integral is only three dimensional and contains a single $\delta$-function that depends linearly on the integration variables.} 

\mb{ 
Eq.~\eqref{eq:col3} certainly holds for weakly correlated systems in the limit of infinite dimensions, but 
may not be applicable 
to  strongly correlated systems. However, if the excitations of the interacting system have a long life time, one might be tempted to  generalize the treatment: replacing the density-of-states $\rho(\epsilon)$ by the spectral density $A(\epsilon)$ which describes the one-particle excitation of the many-body system. 
 In the next section we will provide a more rigorous derivation and show that indeed such a simple substitution works, even if well defined quasiparticle excitations cannot be defined.}


\subsection{Local quantum Boltzmann equation from a DMFT NEGF perspective}
\label{secqbe}

Let us now derive the QBE 
from a NEGF perspective in the DMFT limit. We start by reviewing the definition of the distribution function \mw{$F(\bold k,t,\omega)$}. (As before, we discuss spatially homogeneous states in this section, so that quantities do not depend on $\bold r$). For an in depth discussion of the distribution function and the quantum Boltzmann equation, consider, e.g., Ref.~\onlinecite{kamenev2011field}. In equilibrium, or in any possibly non-equilibrium time-translational invariant state, the spectral function is defined as
\begin{align} 
\label{kaxslcsx.a}
A(\bold k,\omega) = -\frac{1}{\pi}\text{Im} G_{\bold k}^R(\omega),
\end{align}
where 
\begin{align}
G_{\bold k}^R(t,t') = -i\theta(t-t')\langle[c_{\bold k\sigma}(t),c^\dagger_{\bold k\sigma}(t')]_+\rangle
\end{align}
is the retarded Green's function. In a time-translational invariant state, all two-time correlation functions depend only on $t-t'$, and the Fourier transform is taken with respect to this time-difference. Furthermore, the occupied  density of states (unoccupied density of states) is given by the corresponding Fourier transform of the hole propagator $G_{\bold k}^<$ (electron propagator $G_{\bold k}^>$),
\begin{align}
G_{\bold k}^<(t,t') &= i\langle c^\dagger_{\bold k\sigma}(t')c_{\bold k\sigma}(t)\rangle,
\\
G_{\bold k}^>(t,t') &= -i\langle c_{\bold k\sigma}(t) c^\dagger_{\bold k\sigma}(t')\rangle.
\end{align}
In equilibrium, the occupied and unoccupied density of states and the spectrum are related through a variant of the fluctuation dissipation theorem\cite{kamenev2011field}
\begin{align}
G^<_{\bold k}(\omega) &= 2\pi i A_\bold k(\omega) f(\omega),\,\,\,
\\
G^>_{\bold k}(\omega) &= -2\pi i A_\bold k(\omega) f(-\omega).
\end{align}
In a non-equilibrium steady state, one thus defines the distribution function as the ratio 
\begin{align}
F(\bold k,\omega)=\frac{G^<_{\bold k}(\omega)}{2\pi iA(\bold k,\omega)}.
\label{svdjahxks}
\end{align} 

The QBE provides an equation of motion for the time-dependent generalization of this energy distribution function. In a general time-evolving state, Eq.~\eqref{svdjahxks} can be generalized to the ansatz\cite{kamenev2011field}
\begin{align}
\label{ljsblkcsa}
G_{\bold k}^<(t,t') = [F_{\bold k} \circ G^A_{\bold k} - G^R_{\bold k} \circ F_{\bold k}](t,t'),
\end{align}
where $F_{\bold k}(t,t')$ depends on two times, and $[C\circ B](t,t') = \int \mathrm d\bar t C(t,\bar t)B(\bar t,t')$ is the real-time convolution. \mw{ The quantity $G^A_{\bold k}(t,t')$ is called advanced Green's function and reads 
\begin{equation}
G_{\bold k}^A(t,t') = +i\theta(t'-t)\langle[c_{\bold k\sigma}(t),c^\dagger_{\bold k\sigma}(t')]_+\rangle \textrm{ .}
\end{equation}}
 The derivation of the Boltzmann equation is based on a separation of timescales: For every two-time quantity  $C(t,t')$ one can introduce the Wigner transform,
\begin{align}
C(t,\omega)=\int \mathrm d s \,e^{i\omega s} C(t+s/2,t-s/2).
\end{align}
The description of the dynamics can be simplified if the dynamics with respect to average time $t$ changes on a timescale $\Delta t$ which is slow compared to the dependence of $C(t_1,t_2)$ on relative time $\new{s=}t_1-t_2$, i.e, slow compared to $1/\Delta\omega$, where $\Delta \omega$ is the scale on which $C(t,\omega)$ varies in $\omega$ \new{(for a discussion of the relative and absolute time scales $s$ and $t$ cf.~below)}. In mathematical terms, the Wigner transform of the convolution can be represented by the Moyal product
\begin{align}
\label{kasksm}
[C\circ B](t,\omega)
=
e^{-\frac{i}{2}[\partial_t^C\partial_\omega^B-\partial_t^B\partial_\omega^C]} C(t,\omega) B(t,\omega),
\end{align}
and the separation of timescales implies that the terms in the Taylor expansion of the exponential are controlled by the small parameter $\Delta t\Delta \omega\ll1$ (gradient expansion). Keeping only the leading (zeroth) order, Eq.~\eqref{ljsblkcsa} becomes\mw{
\begin{align}
\label{ksbkcql}
G_{\bold k}^<(t,\omega) &= F_{\bold k}(t,\omega) [G^A_{\bold k}(t,\omega) - G^R_{\bold k}(t,\omega)] 
\\&
= 2\pi i A_{\bold k}(t,\omega)F_{\bold k}(t,\omega),
\label{ksbkcql01}
\end{align}}
where we used that the advanced Green's function is given by \mw{$G^A_{\bold k}(t,\omega)=G^R_{\bold k}(t,\omega)^*$}. By analogy to Eq.~\eqref{svdjahxks},  Eq.~\eqref{ksbkcql01} explains the interpretation of \mw{$F_{\bold k}(t,\omega)$} as the distribution in a time-dependent state. 

To derive an equation of motion for the distribution function one can use the Dyson equation 
\begin{align}
[(G_{\bold k,0})^{-1} - \Sigma_{\bold k}]\circ G_{\bold k}=1
\end{align} 
on the Keldysh contour to get\cite{kamenev2011field}
\begin{align}
\label{axknlqsk;a}
(G^R_{\bold k,0})^{-1} \circ F_{\bold k}
-
 F_{\bold k} \circ (G^A_{\bold k,0})^{-1}
= 
\Sigma_{\bold k}^< +  \Sigma_{\bold k}^R\circ F_{\bold k}  - F_{\bold k} \circ \Sigma^A_{\bold k}.
\end{align}
Within  DMFT, the lattice problem is mapped to an impurity model, where the environment is constructed such that the impurity Green's function and self-energy equal the local lattice Green's function and self-energy. (For an introduction to non-equilibrium DMFT, see Ref.~\onlinecite{Aoki2014}.) The impurity Green's functions (which are local by definition, and carry no $\bold k$ index) satisfy the Dyson equation $[(
\new{{\cal G}_0})^{-1} - \Sigma]\circ G=1$, where the noninteracting propagator (Weiss field) is given by $
\new{{\cal G}_0} = [i\partial_t - \mu -\Delta]^{-1}$, with the hybridization function $\Delta$. We thus write Eq.~\eqref{axknlqsk;a} for the DMFT impurity model, where only local ($\bold k$ independent) quantities  enter,
\begin{align}
\label{kwdqbodqwlw}
i(\partial_t + \partial_{t'}) F
&= 
\Big[
\Sigma^< +  \Sigma^R\circ F  - F \circ \Sigma^A
\Big]
\\
\nonumber
&+
\Big[
\Delta^< +  \Delta^R\circ F  - F \circ \Delta^A
\Big].
\end{align}
This equation is still exact \new{(in the DMFT limit of infinite dimensions)}. Via the DMFT self-consistency and the local self-energy diagrams in the impurity model, $\Delta$ and $\Sigma$ are non-linear functionals of the Green's functions, i.e., functionals of the local Green's function $G^R(t,\omega)$ (or spectral function) and the local distribution $F(t,\omega)$.

To derive a QBE, we make two approximations: First, we assume the separation of timescales between the relative-time dependence and the average-time dependence of spectral and distribution functions. In the excited Mott phase, spectra and distribution functions can be assumed to be relatively smooth in $\omega$-space, so that the relative-time evolution happens on a timescale of the inverse bandwidth. 
\me{The average time can be associated, on the other hand,  with thermalization and impact ionization processes, which lead to a change of the distribution function.} 
\me{In actual simulations, one may verify that the average-time dynamics of the distribution function is slow compared to the relative time dynamics, and thus {\em a posteriori} justify that a separation of timescales is a reasonable approximation. In the present case, we will see that the spectral functions are relatively featureless over the full bandwidth, such that the timescale for the relative-time dynamics is set by the inverse bandwidth, while processes like thermalization and impact ionization are slower by at least an order of magnitude. Even for fast impact ionization processes the QBE  works surprisingly well.
 }

To implement this approximation, one simply keeps only the leading (zeroth) order term in the expansion \eqref{kasksm} of the Wigner transform of the convolutions in the right-hand side of Eq.~\eqref{kwdqbodqwlw}:
\me{
\begin{align}
\partial_t F&(t,\omega)
= 
-i\Big[
\Sigma^<(t,\omega) + F(t,\omega) [\Sigma^R(t,\omega)-\Sigma^A(t,\omega)] 
\Big]
\nonumber
\\
&+
\Big[
\Delta^<(t,\omega) + F(t,\omega) [\Delta^R(t,\omega)-\Delta^A(t,\omega)] 
\Big].
\label{kwdqbodqwlw01}
\end{align}
}

In the same spirit, the leading approximation to Eq.~\eqref{kasksm} will be used in the evaluation of $\Sigma$ and $\Delta$ in terms of the Green's function. 
The right hand side in Eq.~\eqref{kwdqbodqwlw01} thereby becomes a functional of the spectrum \mw{$A(t,\omega)=[G^A(t,\omega)-G^R(t,\omega)]/(2\pi i)$ and the distribution function $F(t,\omega)$. } 
In principle, this should be supplemented by a second equation for the spectral function $A(t,\omega)$ itself. However,  in many cases it is observed that even for a strongly correlated system, where $G^R$ is entirely different from the noninteracting $G^R_0$, the spectrum remains relatively rigid in the dynamics. We therefore attempt to close 
Eq.~\eqref{kwdqbodqwlw01} by assuming a rigid density of states, \mw{
\begin{align}
\label{gsgsjdkedkds}
G^R(t,t') = -i \theta (t-t')\int d\omega A(\omega)e^{-i\omega (t-t')},
\end{align}}
where \mw{$A(\omega)$} is the equilibrium spectral function. With a given input \mw{$A(\omega)$}, Eq.~\eqref{kwdqbodqwlw01} is an equation for the distribution function alone.

The exact functional form of $\Sigma(t,\omega)$ in terms of $F(t,\omega)$ and the spectrum \mw{$A(\omega)$} is complicated, because it includes high-order diagrams. The general self-energy diagram on the Keldysh contour has the form
\begin{align}
\Sigma(t,t')
=
U 
G(t,t_1)
G(t,t_2)
G(t_3,t)
\Gamma(t_1,t_2,t_3,t')
\end{align}
with some two-particle irreducible vertex $\Gamma(t_1,t_2,t_3,t')$. To arrive at a QBE with a generic structure, we assume that, though being renormalized with respect to the bare interaction,  $\Gamma$ is fairly local in time (at least in the Mott phase), so that for the sake of evaluating the scattering term of the Boltzmann equation  we can replace it by a renormalized $\tilde U$. The latter becomes the only adjustable parameter in the comparison to non-equilibrium DMFT, just like the parameter $\alpha$ in the local QPBE \eqref{eq:col3}.

With this, the self-energy in \eqref{kwdqbodqwlw01} is replaced by
\begin{align}
\Sigma(t,t')
=
\tilde U
U
G(t,t')
G(t,t')
G(t',t).
\label{msbkskcbswx}
\end{align}
Writing the equation for the $G^>$ and $G^<$ components of $G$, and using the rigid density of states \eqref{gsgsjdkedkds}  in combination with the leading \mw{order} approximation to the Moyal product, i.e, \mw{$G^>(t,\omega)=-2\pi i A(\omega) [1-F(t,\omega)]$ and $G^<(t,\omega)=2\pi i A(\omega) F(t,\omega)$}, the scattering term (involving $\Sigma$) in Eq.~\eqref{kwdqbodqwlw01} becomes \mw{
\begin{align}
\me{2\pi}
U\tilde U
\! \!\int   \mathrm d  \omega_1 \mathrm d  \omega_2 &\mathrm d  \omega_3 \delta \left ( \omega_0\! +\! \omega_1\! -\! \omega_2 \!-\! \omega_3  \right ) A(\omega_1) A(\omega_2) A(\omega_3) 
\nonumber\\
& \times \mathcal{P}[F(\omega_0),F(\omega_1),F(\omega_2),F(\omega_3)],
\label{knfff}
 \end{align}}
 with the phase-space factor \eqref{gdjadsljbja}.
\me{
We note that this scattering term of the quasi-particle Boltzmann equation yields 
$\frac{\partial F (\omega)}{\partial t }  = 0$ if $F$ is the (equilibrium) Fermi function, and in turn, the Fermi function is the only time-invariant fixed point if detailed balance is assumed, i.e., if the integrand in Eq.~\eqref{knfff} vanishes at every point.}

Depending on the non-linear structure of the DMFT self-consistency, a term of analogous functional form can emerge from the second ``scattering term''  $\Delta^<(t,\omega) + F(t,\omega) [\Delta^R(t,\omega)-\Delta^A(t,\omega)]$ in Eq.~\eqref{kwdqbodqwlw01}. This could be absorbed into a further renormalization of the scattering matrix element $\tilde UU$. (However, we note that for the closed form self-consistency for a Bethe lattice, $\Delta(t,t')=G(t,t')$, the term vanishes after making the zeroth order gradient 
approximation 
in the Moyal product Eq.~(\ref{kasksm}), i.e., \mw{$G^<(t,\omega)=2\pi i A(\omega) F(t,\omega)$}.)

In conclusion, the above argument suggests that in the presence of a rigid density of states the evolution of the distribution function is given by a local QBE of the same structure as Eq.~\eqref{eq:col3},  \mw{
\begin{align}
&\frac{\partial F (\epsilon_0 )}{\partial t }  =
\alpha\! \!\int   \mathrm d  \epsilon_1 \mathrm d  \epsilon_2 \mathrm d  \epsilon_3 A(\epsilon_1) A(\epsilon_2) A(\epsilon_3) \,\,\,\times
\nonumber\\
& \times  \delta \left ( \epsilon_0\! +\! \epsilon_1\! -\! \epsilon_2 \!-\! \epsilon_3  \right ) \mathcal{P}[F(\epsilon_0),F(\epsilon_1),F(\epsilon_2),F(\epsilon_3)],
 \label{eq:col444.e,d}
\end{align}}
where $\alpha$ is an unknown free parameter and \mw{$A(\epsilon)$} the interacting spectral function in equilibrium which is 
assumed 
to remain unchanged under the non-equilibrium dynamics. 


\subsection{Laser excitations} \label{chap:laser}

With the electron-electron collision term we can simulate how an out-of-equilibrium distribution relaxes back to a Fermi-Dirac distribution. In principle, one could take the distribution function obtained from the short-time evolution in DMFT, i.e. after a laser excitation, as an initial state of the Boltzmann equation. Here we instead supplement the Boltzmann equation with a natural extension that does also reproduce the transfer of occupied weight from the lower to the upper band. We use two different ways to implement the spectral weight transfer, motivated by Fermi's golden rule and time-dependent perturbation theory applied to electrons in a given band.

\subsubsection{Fermi's golden rule} \label{chap:fermi}

In a given band structure, the excitation of an electron due to a single-photon process can be regarded as a scattering process of an electron with a photon, and we can treat it within the Boltzmann framework as an additional collision term. 
We do not describe the reaction of the laser field to the excitation and we assume only a single laser frequency $\Omega$ which leads to the simple laser collision term \mw{
\begin{equation}
\begin{split}
\left ( \frac{\partial F (\epsilon_0 )}{\partial t } \right ) _\textrm{col-laser}  =& \int \mathrm d  \epsilon_1 A(\epsilon_1) W_\textrm{laser} \left ( \epsilon_0, \epsilon_1 \right )   \\
&    \times \Big [-  F(\epsilon_0)  (1-F(\epsilon_1)) \\
 &    +   (1-F(\epsilon_0))  F(\epsilon_1) \Big ] \textrm{ ,}
\end{split} \label{eq:col-laser}
\end{equation}}
with the transition amplitude
\begin{equation}
W_\textrm{laser} \left ( \epsilon_0, \epsilon_1 \right ) =  I \left [ \delta \left ( \epsilon_0 - \epsilon_1 - \Omega  \right ) + \delta \left ( \epsilon_0 - \epsilon_1 + \Omega  \right ) \right ]  \label{eq:col-laser-trans}
\end{equation}
according to Fermi's golden rule.\cite{Dirac1927, Fermi1950} It includes stimulated emission as well as absorption. The quantity $I$ is proportional to the intensity of the radiation \mw{ and the transition matrix element between the quantum mechanical states. Here we assume that the transition matrix element is the same for all transitions, i.e. a scalar incorporated in $I$.} In order to describe laser pulses with some time-envelope we use 
Eq.~\eqref{eq:col-laser-trans}
with a time-dependent pre-factor $I(t)$ in the transition amplitude. 
This approximation is well-justified if the envelope function of the laser pulse changes slowly compared to the period of the laser-field ($\tau = \frac{2 \pi}{\Omega}$). 

\subsubsection{First order perturbation theory} \label{chap:pert}
The argument that the laser period is much 
\new{shorter} 
than the time-envelope function is not strictly fulfilled for the case we are going to study. Therefore we also apply first order time-dependent perturbation theory in order to 
estimate the
transition rate. We will consider a laser field of the form 
\begin{equation}
\bold E(t) = \bold n E(t) = \bold n \underbrace{E_0 e ^{- {\left ( \frac{t-t_0}{\sigma} \right ) }^2 }}_{\equiv \sqrt{I(t)}} \sin(\Omega (t-t_0) ) \label{eq:efield}
\end{equation}
with the unit vector $\bold n$. We assume that it couples via the operator $-\bold{\hat r}$ so that the perturbing potential reads $\hat V(t) = E(t) \hat O$ with $\hat O \equiv - \bold {\hat r} \cdot \bold n$. 

For the perturbing potential $\hat V(t)$ the transition probability to find an electron that initially was in the state $\ket{0}$ at $t=0$ in state $\ket{1}$ at time $t$ is 
\begin{equation}
p_{01}(\epsilon_0 , \epsilon_1 , t) =   \Big | \bra{1} \hat O \ket{0}   \int_0 ^ t \mathrm d \tau E(\tau) e^{i (\epsilon_1 - \epsilon_0) \tau} \Big | ^2 \label{eq:p01}
\end{equation}
with the energies $\epsilon_0$ ($\epsilon_1$) of the initial (final) state ($\hbar =1$). 
In the following we assume for simplicity that the transition matrix element is one, i.e. $\bra{1} \hat O \ket{0} = 1$ for all states\mw{.
In analogy with the derivation of Fermi's golden rule we} replace the transition rate in Eq.~\eqref{eq:col-laser} by
\begin{equation}
W_\textrm{laser} \left ( \epsilon_0, \epsilon_1  , t \right ) = \frac{\partial}{\partial t} p_{01}(\epsilon_0 , \epsilon_1 , t) \textrm{ .} \label{eq:lasertens}
\end{equation}
Let us emphasize that the transition amplitude $W_\textrm{laser}$ is now time dependent and can become negative as well. 
The latter 
represents the coherent dynamics of electrons which are brought back to their original state after being excited. Such coherent processes are usually excluded in Boltzmann theory as they average to zero over longer timescales. On short timescales they can 
produce internal inconsistencies, at least in principle, since they are associated with negative transition probabilities, which  could lead to negative populations.  However, if the electron-electron scattering is much slower than the frequency of the coherent processes we can still employ Eq.~\eqref{eq:lasertens} as an approximation since it is a valid (first-order) description if we only consider the (phase coherent) laser excitation between the states $\epsilon_0$ and $\epsilon_1$. 

\subsection{\mb{Spectral density for QBE}} \label{chap:dos}

As described in the previous sections, our quantum Boltzmann equation is based on the assumption of a relatively rigid local density of states \mw{$A(\omega)$}, with the Boltzmann equation describing the distribution function $F(\epsilon)$ with respect to this rigid density of states.
This approximation is justified retrospectively by comparison to \mw{non-equilibrium} DMFT results, where it is found to work under certain conditions.  
For our calculations we use the equilibrium spectral density $A(\epsilon)$ obtained from DMFT calculations for a certain chemical potential $\mu_0$ and temperature $T_0$ and assume that its structure remains unchanged even when energy is pumped into the system, i.e. \mw{
\begin{align} 
\label{eq:dosDef}
A(\omega) = -\frac{1}{\pi}\text{Im} G^R(\omega) \quad \forall t,
\end{align}
where $G^R(\omega)$ is the local retarded Green's function of a system with $\mu_0$ and $T_0$.}

\begin{figure}
 \includegraphics[width=8.5cm]{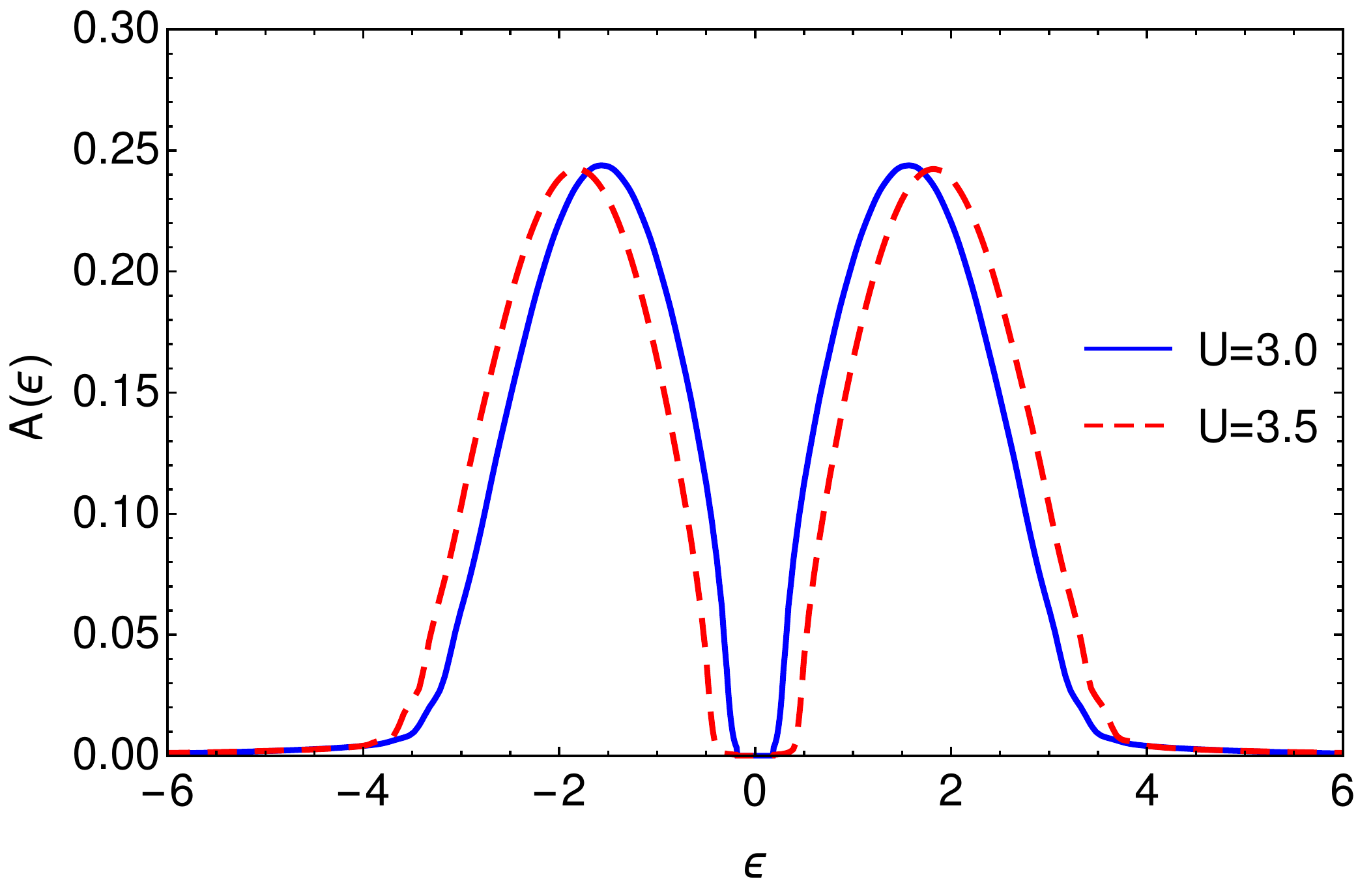}
 \caption{Spectral functions obtained from DMFT calculations for a hypercubic lattice at half-filling ($\mu = 0$), inverse temperature $\beta=5$ and two different Hubbard-interactions $U$. For both interactions the system is in the Mott-insulating phase. (taken from Ref. \onlinecite{Werner2014})
}\label{fig:specdens}
\end{figure}

The density of states 
used in the simulations corresponds to a Mott insulator with Hubbard bands and is shown in Fig.~\ref{fig:specdens}.
Here,  quasiparticles are not particularly well defined since
the equilibrium DMFT self-energy \cite{Werner2014}
is 0.4 at the upper edge of the upper Hubbard band and -- for generating
the Mott gap --  even larger (larger than the bandwidth) at its
lower edge.\footnote{The DMFT self-energy in the Mott-Hubbard bands remains finite
also in the $U\rightarrow \infty$ limit, albeit it becomes slightly smaller and symmetric.}
Nonetheless the 
Boltzmann equation
is found to 
produce meaningful results
which demonstrates that the local Boltzmann equation does not have to be built on a quasiparticle approximation, as explained in Sec.~\ref{secqbe}.

%

\subsection{Numerical implementation} \label{chap:numImp}

In principle we could solve \mw{Eq.~\eqref{eq:col444.e,d}} on a given $\epsilon$-grid, using a 
Runge-Kutta scheme with direct numerical integration of the scattering term at runtime. Numerically, it turns out to be advantageous to project the problem onto an orthonormal basis $\Phi_i(\epsilon)$. In this work we 
employ 
orthonormal discontinuous piece-wise polynomial functions up to the 2-nd polynomial order as basis functions \mw{(see appendix \ref{chap:appBasis})}.
Using the expansion coefficients of these basis functions, \mw{
\begin{subequations}
\begin{align}
a_i (t) & \equiv \int \mathrm d \epsilon \Phi_i (\epsilon) F(t,\epsilon) \textrm{ ,} \\
n_i & \equiv \int \mathrm d \epsilon \Phi_i (\epsilon),
\end{align}
\end{subequations}}
\mw{Eq.~\eqref{eq:col444.e,d}} becomes
\begin{equation}
\begin{split}
\frac{\partial a_i(t)}{\partial t} &= \alpha \sum_{j,k,m,n} \mathbb S_{ijkmn}  \\
& \times\Big (- a_j (t) a_k (t) \left ( n_m - a_m (t)  \right ) \left ( n_n - a_n (t)  \right ) \\
 & +\left ( n_j - a_j (t)  \right )  \left ( n_k - a_k (t)  \right ) a_m (t) a_n (t) \Big ),
\end{split} \label{eq:col4}
\end{equation}
with scattering amplitude parameter $\alpha$ and normalized scattering-tensor \mw{
\begin{equation}
\begin{split}
\mathbb S_{ijkmn} =&  \int \mathrm d \epsilon _0 \mathrm d \epsilon _1 \mathrm d \epsilon _2 \mathrm d \epsilon _3  \Phi_i (\epsilon_0) \Phi_j (\epsilon _0) \Phi_k (\epsilon _1) \Phi_m (\epsilon _2) \Phi_n (\epsilon _3)  \\ 
 &\times A (\epsilon_1) A (\epsilon_2) A (\epsilon_3)  \delta \left ( \epsilon_0 + \epsilon_1 - \epsilon_2 - \epsilon_3 \right ) \textrm{ .}
\end{split} \label{eq:scat} 
\end{equation}}
Note that $\mathbb S_{ijkmn}$ contains all the information about the system i.e.,  the density-of-states, and is independent of the distribution function and time. 
In our numerical approach we pre-calculate and store the tensor elements. 

When the scattering-tensor is known, the time-propagation of Eq.~\eqref{eq:col4} for a given initial distribution $a_i(t_0)$ is numerically cheap as the evaluation of the right-hand side only consists of tensor-vector multiplications. 

\mw{The numerical implementation of the Fermi's golden rule laser excitation (section \ref{chap:fermi}) is done in the same way as for the electron-electron collision term, where the time-dependent part can be pulled out of the integral. The corresponding laser-scattering tensor reads
\begin{equation}
\begin{split}
\mathbb S _{ijk} ^{\textrm{laser}} = \int & \mathrm d \epsilon_0 \mathrm d \epsilon_1  A(\epsilon_1) \Phi_i(\epsilon_0) \Phi_j(\epsilon_0) \Phi_k(\epsilon_1) \\
&\times \left [ \delta \left ( \epsilon_0 - \epsilon_1 - \Omega  \right ) + \delta \left ( \epsilon_0 - \epsilon_1 + \Omega  \right ) \right ] \textrm{ .} 
\end{split}
\end{equation}
In case of laser excitations directly calculated from perturbation theory (section \ref{chap:pert}) the transition amplitude does depend on time explicitly which means that we have to calculate a laser-scattering tensor for every time step.  
}

\section{Results and discussion} \label{chap:results}

\subsection{Two time-scale relaxation dynamics} \label{chap:twotimescales}

In this section we compare the dynamics obtained by the 
\mb{QBE} to the non-equilibrium DMFT result for a hypercubic lattice at half-filling. 
The noninteracting density of states is $\rho_0(\epsilon)=\exp(-\epsilon^2/W^2)/\sqrt{\pi}W$ and we use $W=1$ as the unit of energy ($1/W$ as the unit of time).
 We discuss Mott insulating systems with Hubbard-interactions $U=3.0$ and $U=3.5$ and set the initial inverse temperature to $\beta=5$. 
 
The 
excitations in the upper Hubbard band can be interpreted as double-occupancies of lattice-sites (doublons) while excitations in the lower Hubbard-band are empty sites (holons). \new{We hence define the  total doublon density $d(t)$ as} 
\mw{
\begin{equation}
d(t) = \int _{0}^\infty \mathrm d \epsilon  F(t,\epsilon)  A(\epsilon) 
\end{equation}}
in the Boltzmann approach.\footnote{\new{Note that this does not include virtual doublon excitations which are present in the Mott insulator even at zero temperature where  $F(t,\epsilon)=0$ for $\epsilon>0$. If one counts both, up- and down-spin, spectral function there is also a factor of two compared to the usual double occupation because each doublon gives a peak in both, spin-up and -down, spectral functions. Since we discuss ratios of double occupations this factor two cancels anyhow.}}

As in Ref.~\onlinecite{Werner2014} we excite the system with laser-pulses at different frequencies $\Omega$ and let it time-propagate until it is thermalized, i.e. until it has reached a Fermi-Dirac distribution again. 
We use a Gaussian time envelope centered at $t_0=6$ for the laser pulse as defined in Eq.~\eqref{eq:efield}, and a pulse width  $\sigma = \sqrt{6}$. 
The strength of the laser pulse is adjusted such that the photo-induced doublon density at 
a given time $\tilde t$
right after the pulse is 0.01, i.e. $D( \tilde t) \equiv d(\tilde t) - d(0) = 0.01$. 
For
the two different laser implementations a different strength of the laser pulse is needed to produce the same number of photo-doped doublons.
We use $\tilde t=15$ for the results given in Tab.~\ref{tab:times}, and $\tilde t=12$ otherwise, in order to directly compare with time dependent data provided in Ref.~\onlinecite{Werner2014}.

\begin{figure}
 \includegraphics[width=8.5cm]{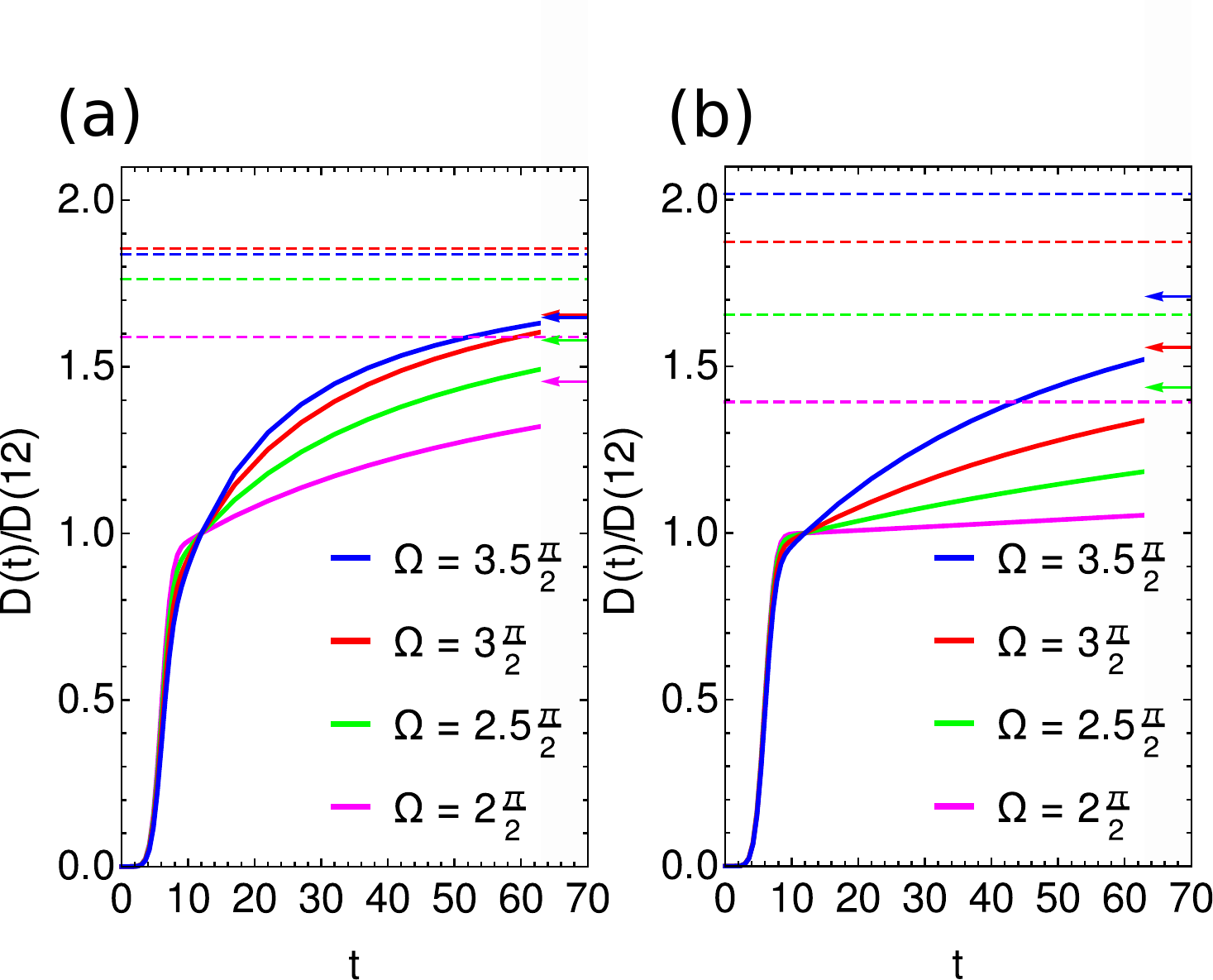}
 \caption{
 Normalized doublon density $D(t)/D(12)$ as a function of time for different laser frequencies (a) $U=3.0$ and (b) $U=3.5$. The dashed lines indicate the final value when the system has reached the new thermal equilibrium; the arrows give the final doublon density for a single exponential function fitted within the time interval $t \in [12,60]$. Note that in a) the blue and red 
 arrows lie almost on top of each other. In (b) there is no pink arrow as the fit was not possible within reasonable tolerance. Here Fermi's golden rule (Section \ref{chap:fermi}) was used for the laser excitation.
}\label{fig:doublon_dens}
\end{figure}

Figure~\ref{fig:doublon_dens} shows that after the laser pulse has created a non-Fermi-Dirac population, the doublon density further increases until the system reaches its new equilibrium (marked by dashed lines). This means that during the thermalization process new doublons (and holons) have to be generated, hence electrons have to be excited across the Mott gap. 
The doublon-holon creation results from 
two different mechanisms: i) impact ionization and ii) multiple-scattering events. Case i) means that a doublon with an initial kinetic energy larger than the gap lowers its energy and excites another electron across the band-gap which generates one doublon and one holon.\cite{Werner2014,Sorantin2018} This process has been shown to be beneficial to the efficiency of correlated solar cells.\cite{Manousakis2010,Assmann2013} The second thermalization process ii) means that a low energy doublon gains kinetic energy through several scatterings with other doublons and holons until its kinetic energy exceeds the gap size. Then it generates another doublon and a holon by lowering its kinetic energy. 
Excited holons undergo analogous processes. 
Within the 
Boltzmann description
these are the only two processes that can lead to the generation of additional doublons. There are further processes in non-equilibrium DMFT 
related to the change of the spectral function.
In particular, spectral weight is filled into the Mott-gap, similarly as upon increasing the temperature.\cite{Mo2004}  

The two mechanisms i) and ii) take place on different timescales which can be seen if we try to fit 
a single-exponential decay \mw{
$D_\textrm{1}(t) = a + b \, \mathrm{exp}[- t / \tau]$} to the doublon density 
$d(t)$
in a short time interval right after the pulse, i.e. for $t \in [12,60]$. If there was only one thermalization mechanism the doublon curve would roughly follow the fitted function over the whole time range. However, this is not the case as can be seen in Fig.~\ref{fig:doublon_dens}. The final value of the doublon density (dashed lines in Fig. \ref{fig:doublon_dens}) deviates significantly from the final value of the fitting function (arrows in Fig. \ref{fig:doublon_dens}). In accordance with Ref.~\onlinecite{Werner2014} we find that the whole time evolution of the doublon density can be described by 
the sum of two exponential 
functions \mw{$D_\textrm{2}(t) = a + b \, \mathrm{exp}[- t / \tau] + c \, \mathrm{exp}[- t / \gamma]$.} The two fitting parameters $\tau$ and $\gamma$ represent the timescales on which the two different thermalization mechanisms described above take place. \mw{Here $\gamma$ corresponds to the short timescale associated with impact ionization and $\tau$ to the long timescale associated with multiple scattering events}.

The two-time relaxation is already qualitatively consistent with DMFT. For a quantitative comparison, we note that the overall timescale for the evolution of the Boltzmann equation \mw{Eq.~\eqref{eq:col444.e,d} 
} is set by the constant $\alpha$. As discussed above, $\alpha$ is treated as an adjustable parameter, as its ab-initio determination is difficult. Nevertheless, we can perform a non-trivial quantitative comparison between the Boltzmann approach and DMFT, by comparing the ratio between different timescales. 

\mb{We choose integer-valued $\alpha$ for each value of $U$ (independent of the laser frequency), such that the short timescale $\gamma$ 
extracted from the 
fit roughly coincides with the short timescale $\gamma_\textrm{DMFT}$ of Ref.~\onlinecite{Werner2014}. The strategy of not fitting the timescales obtained from Ref.~\onlinecite{Werner2014} more precisely, is motivated by the fact that the choice of $\tilde t$ and the fitting time range have a sizeable impact on the value of the time constants (see a more precise discussion below). This makes it meaningless to fit $\alpha$ with more than one significant digit.
Interestingly, in spite of the limitations due to the fitting procedure, we clearly}  find that the second (longer) time scale $\tau$ shows the same order of magnitude as the DMFT result $\tau_\textrm{DMFT}$ (see Tab.~\ref{tab:times}), and the ratio $\frac{\tau}{\gamma}$ is almost independent of $\alpha$ for both DMFT and the Boltzmann approach.

\begin{table}
\centering
\begin{tabular}{c| c ||c|c|c||c|c|c|c | c|c|c|}
\cline{3-12}
\multicolumn{2}{c|}{} &  \multicolumn{3}{|c||}{DMFT } & \multicolumn{7}{|c|}{\mb{ QBE 
}} \\
\hline
 \multicolumn{1}{|c|}{U} 	&	$\Omega$	&	$\gamma$	&	$\tau$	& 	$\tau/\gamma$ & $\gamma$	&	$\tau$	& 	$\tau/\gamma$   & \it{$\gamma$}	&	\it{$\tau$}	& 	\it{$\tau/\gamma$} &  $\alpha$  \\
\hline \hline
 \multicolumn{1}{|c|}{3}	    & 3.5 $\frac{\pi}{2}$  & 13   &   60   &   4.50   &   15   &   95   &   6.48   &\it{11} &\it{70} &\it{6.37} &8\\
 \hline
  \multicolumn{1}{|c|}{3}    & 3 $\frac{\pi}{2}$  & 15  &   61   &   4.09   &   18   &   91   &   5.01  &\it{13}&\it{68} &\it{5.20} &8\\
 \hline 
  \multicolumn{1}{|c|}{3}    & 2.5 $\frac{\pi}{2}$  & 17   &   65   &   3.93   &   22   &   95   &  4.23  &  \it{16}&\it{76}&\it{4.84} &8\\
 \hline \hline
  \multicolumn{1}{|c|}{3.5}	    & 3.5 $\frac{\pi}{2}$  & 44 &  376    &   8.55   &  39   &  231   &   5.88  & \it{ 26}&\it{131}&\it{5.03}&5 \\
 \hline  
  \multicolumn{1}{|c|}{3.5}	    & 3 $\frac{\pi}{2}$  & 48   &  257   &  5.31   &  53   &  254   &   4.85  &  \it{32} &\it{167} &\it{ 5.18}&5\\
 \hline 
\end{tabular}
\caption{Results for different interaction parameters $U$ and laser frequencies $\Omega$. The non-italic numbers in the \mw{QBE} section are obtained from fits to the doublon density over the whole thermalization time and the italic ones are from fits within the interval $t \in [15,60]$ with fixed final doublon value, i.e. as in Ref. \onlinecite{Werner2014} for non-equilibrium DMFT.} \label{tab:times}
\end{table}

As mentioned above, there is some freedom in the determination of the time constants listed in Tab.~\ref{tab:times}.
First of all,
the times $\gamma$ and $\tau$ are extracted from fits of a 
sum of two exponential 
functions to the doublon curve over the whole time range. 
As can be seen in Fig. \ref{fig:doublon_dens}, impact ionization takes place on the same timescale as the laser excitation \mw{for the case $U=3$}. In other words, while the laser is switched on, the excited doublons immediately start to produce additional doublons through impact ionization.  Since we further normalize the number of doublons  to $0.01$ at time $\tilde t = 12$ (or $\tilde t = 15$), \mb{to be consistent with the approach in Ref.~\onlinecite{Werner2014}}, the details of the dynamics 
depend on the exact timing and shape of the laser as well as on the normalization time $\tilde t$. For $U=3.5$ the laser and impact ionization timescales are well separated which makes the dynamics more independent of the exact laser shape and duration.

\begin{figure}
 \includegraphics[width=8.5cm]{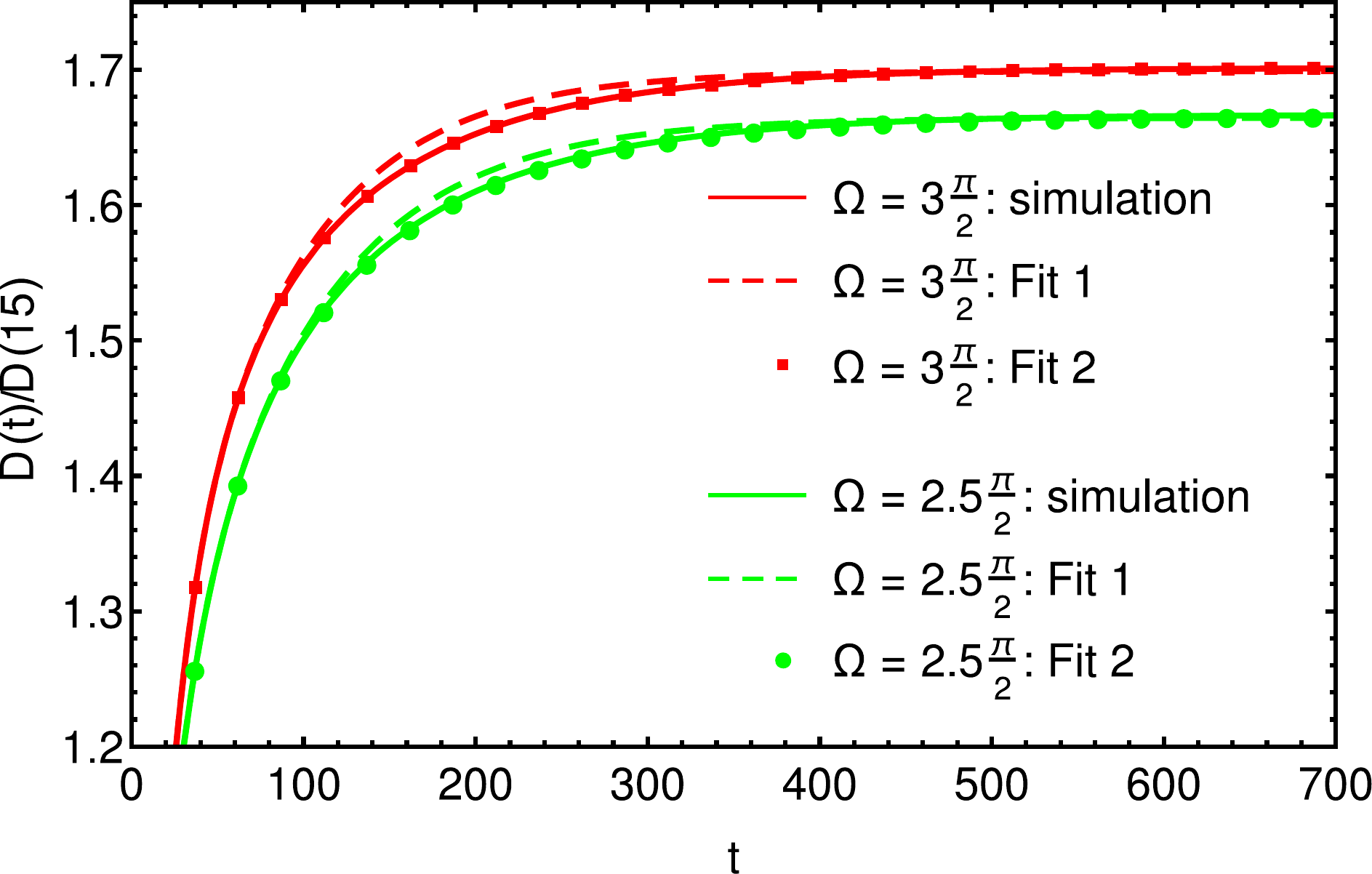}
 \caption{Normalized doublon density as a function of time for $U=3.0$ and different laser frequencies $\Omega$ in a large time interval. The solid lines are the simulated doublon density, the dashed lines represent the double-exponential fit within the time interval $t \in [15,60]$ and with fixed final doublon number. The dots indicate the double-exponential fit obtained from an even larger total simulated time $t \in [15,1700]$. 
}\label{fig:nTU3d0}
\end{figure}

\begin{figure}
 \includegraphics[width=8.5cm]{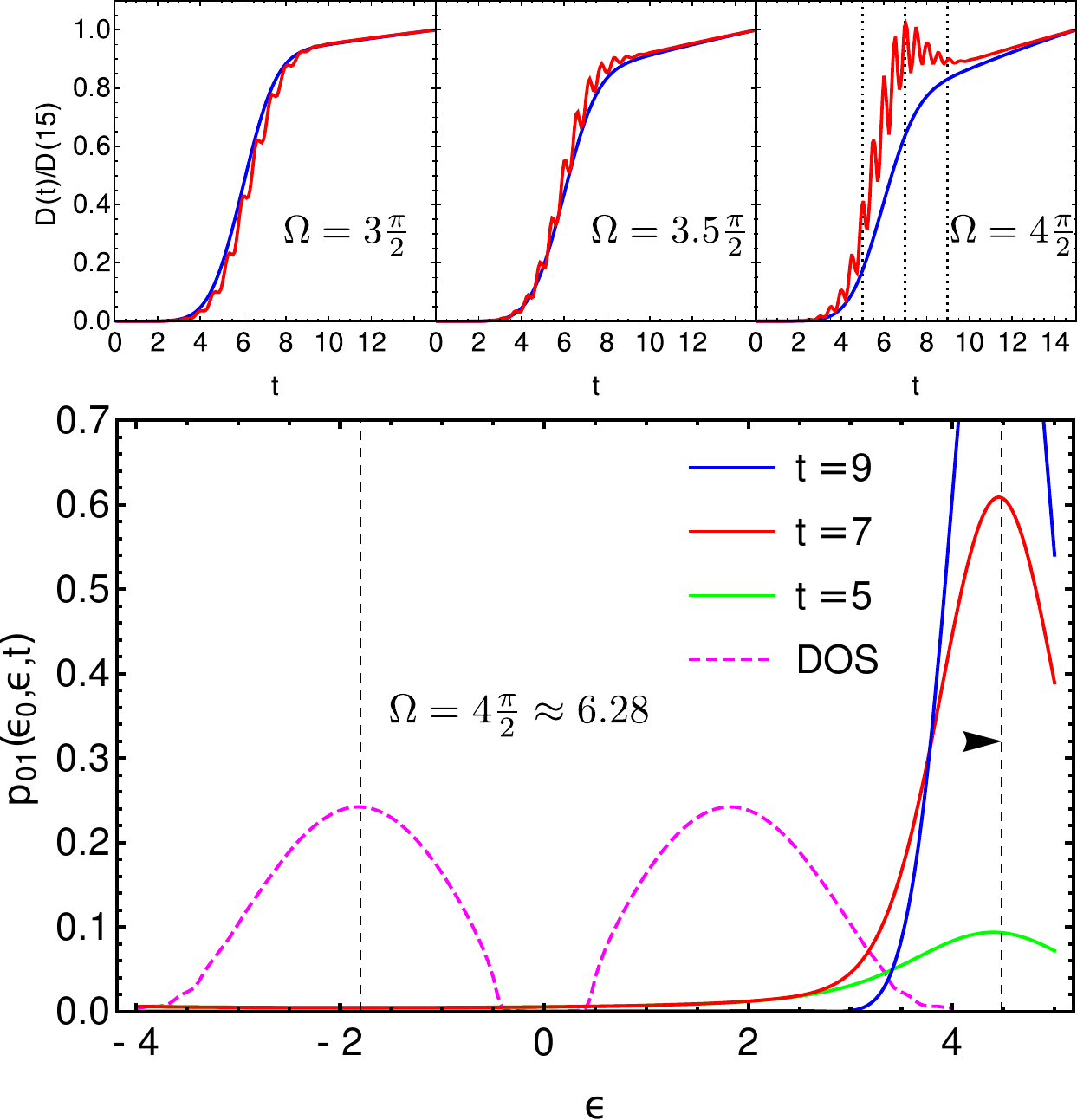}
 \caption{Upper panel: Time-dependent doublon density for \mb{QBE 
 } comparing Fermi's golden rule (blue lines) and first order perturbation theory (red lines) for the laser transition at different frequencies for $U=3.5$. Lower panel: Transition probability $p_{01}$ [Eq.~\eqref{eq:p01}] from a specific initial state energy $\epsilon_0 = -1.8$, at different times (dashed lines in the right upper panel), for $\Omega = 6.28$ and $U=3.5$. For the related non-equilibrium DMFT result see Fig.~2 of \protect Ref.~\onlinecite{Werner2014}. 
 }
\label{fig:laser_exc}
\end{figure}

Furthermore, when comparing the times to the non-equilibrium DMFT results one should keep 
in mind that the non-equilibrium DMFT has only access to short times after the excitation (maximum time $t=60$ in Ref.~\onlinecite{Werner2014}). The final (equilibrium) doublon value was determined from the temperature corresponding to the total energy of the system, and the remaining constants were obtained from a fit in the time interval $t \in [15,60]$.
However, within the Boltzmann framework we can obtain the thermalization times $\tau$ and $\gamma$ from fits of \mw{$D_\textrm{2}(t)$} to the doublon density over the full thermalization time (e.g. $t \in [15,1700]$ for $U=3.0$; \mw{``Fit 2''} in Fig. \ref{fig:nTU3d0}). 
In order to estimate the error arising from the fact that non-equilibrium DMFT has only a limited time interval for the fit we perform a second fit analogous to the DMFT fit. That is,  we assume that the constant $a$ in \mw{$D_\textrm{2}$} is equal to the final doublon value, $a = D(t_\textrm{max})$ and we obtain the other coefficients from fitting within $t \in [15,60]$ (``Fit 1'' in Fig.~\ref{fig:nTU3d0}). All results are collected in Tab.~\ref{tab:times} and the different fits are shown together with the simulated doublon densities in Fig.~\ref{fig:nTU3d0} for two laser frequencies.  Some deviations are visible at intermediate times between $t=100$ and $t=300$.

\begin{figure*}
 \includegraphics[width=18cm]{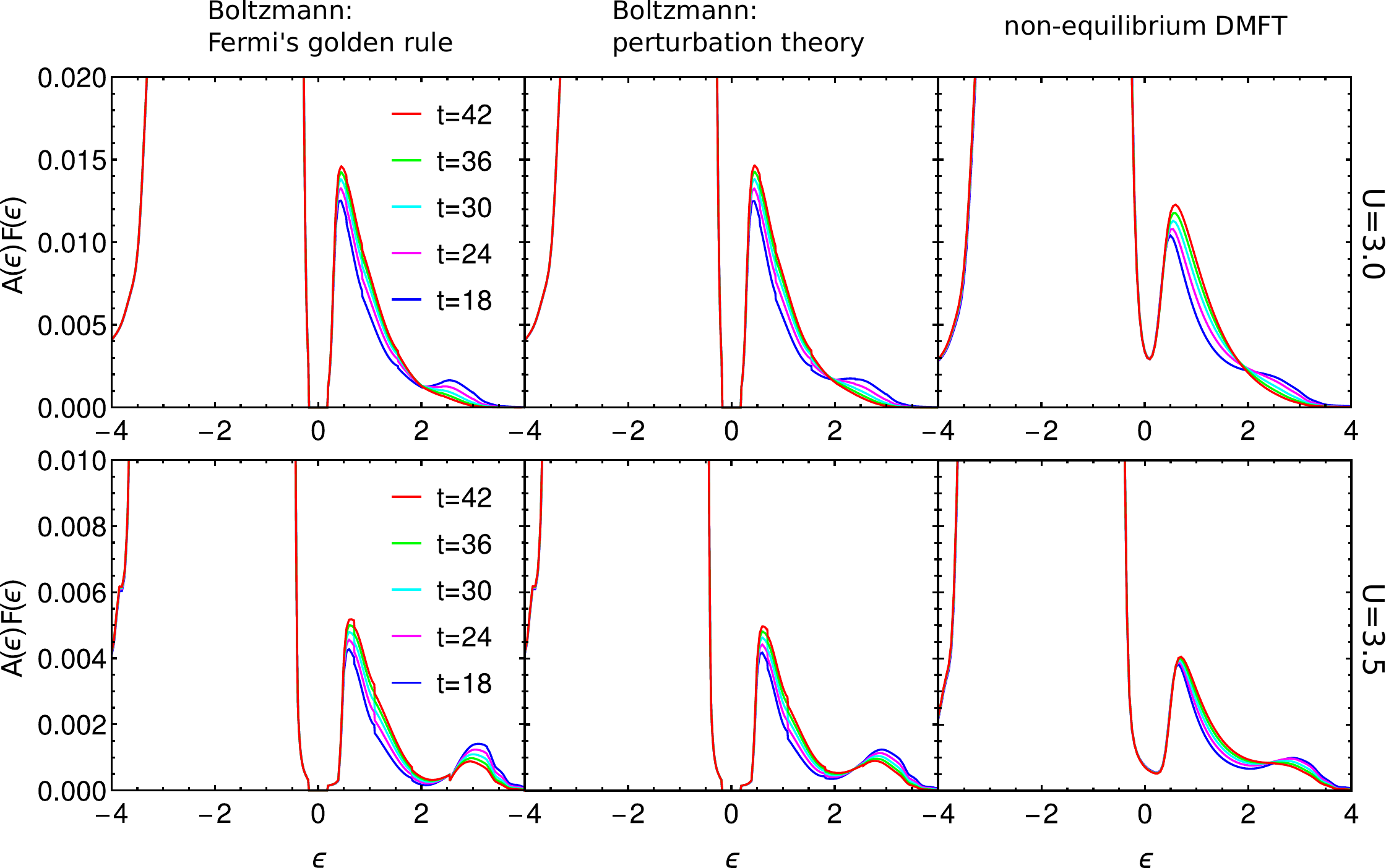}
 \caption{Electron population for $U=3.0$ (first row) and $U=3.5$ (second row) at different times. The figures in the first column are obtained from \mb{QBE 
 } simulations with the Fermi's golden rule laser transition, the figures in the second column are obtained from \mb{QBE 
 } simulations with first-order perturbation theory laser transitions and the figures in the third column show the photo-emission spectra obtained from the lesser component of the DMFT Green's function\cite{Werner2014}.  
 For $U=3.0$ the number of photo-doped doublons after the laser pules is $D(12) = 0.0056$ and the laser frequency is $\Omega = 3.5 \pi /2$, for $U=3.5$ we have $D(12) = 0.0021$ and a laser frequency of $\Omega = 4 \pi /2$ in accordance with Ref. \onlinecite{Werner2014}. \mw{For both $U$-values the initial inverse temperature is $\beta = 5$.} 
}\label{fig:time_distr}
\end{figure*}

\mw{The data in Tab. \ref{tab:times} in addition with the previous discussion shows that the doublon-relaxation timescales of non-equilibrium DMFT and \mb{QBE}
are similar within the numerical and methodological tolerance. }
This result may be unexpected given the fact that the validity of the 
QBE, which assumes a rigid spectrum, is not a priori clear for the description of strongly correlated systems. 

\begin{figure}
 \includegraphics[width=8.5cm]{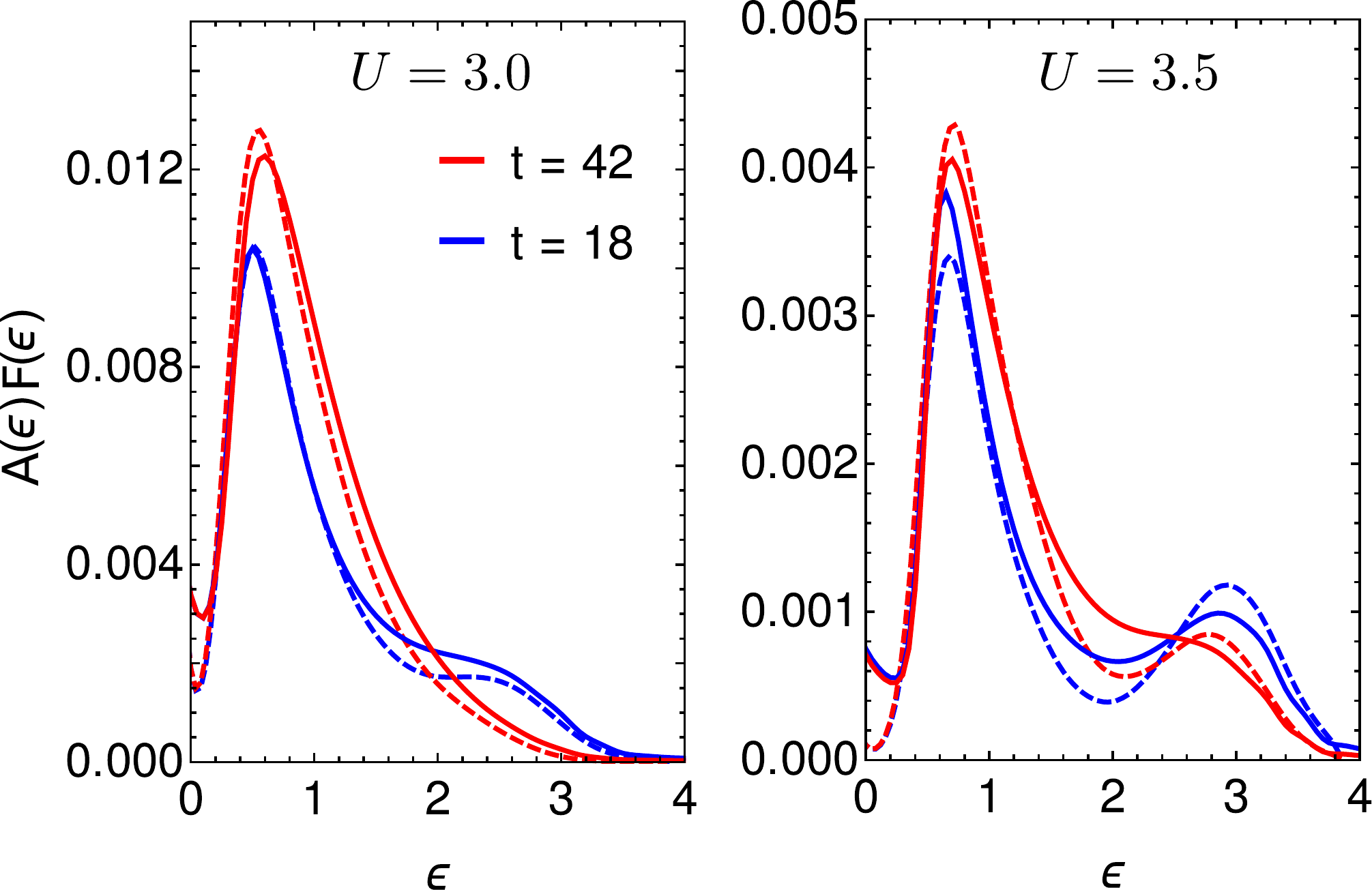}
 \caption{Electron population for $U=3.0$ (left panel) and $U=3.5$ (right panel) from \mb{QBE} 
 with laser transition in perturbation theory (dashed lines) and non-equilibrium DMFT (solid lines). Here, the Boltzmann distributions of the middle column of Fig. \ref{fig:time_distr} were smoothed by convolution with a Gaussian with $\sigma=0.14$ in order to resemble the smoothing inherent in the DMFT data.
}\label{fig:distr_smooth}
\end{figure}

\subsection{Coherent laser excitations} \label{chap:laserEx}
When the first order perturbation theory laser is used to model the absorption process (Section \ref{chap:pert}), there are some differences compared to the simpler implementation with Fermi's golden rule. First, we can observe that the excitation due to the laser field is not monotonic but oscillates with 
twice 
the laser frequency (upper panel Fig. \ref{fig:laser_exc}). For smaller frequencies ($\Omega = 3 \frac{\pi}{2}, 3.5 \frac{\pi}{2}$) the {
doublon density roughly 
follows} 
the 
prediction from Fermi's golden rule,
whereas for higher frequencies ($\Omega = 4 \frac{\pi}{2}$) we observe that the doublon density in the perturbation theory implementation 
first increases more strongly, but then decreases at the end of the laser pulse to approach the same value as given by the Fermi's golden rule implementation.
That is, we have a maximum in $D(t)$ at $t\approx 7$ in  Fig.~\ref{fig:laser_exc} (upper right panel). This behavior gets 
more pronounced
as the frequency increases, as was confirmed by an additional simulation at $\Omega = 4.5 \frac{\pi}{2}$ (not shown). 
A similar 
effect was also observed in non-equilibrium DMFT simulations (compare upper right panel of Fig.~2 of Ref. \onlinecite{Werner2014}) and finite system simulations.\cite{Kauch2018}

With the simpler Boltzmann approach, the physics behind this behavior can be understood.
The laser frequency $\Omega = 4 \frac{\pi}{2}$ is so large that only the outermost regions of the 
density of states 
are connected by direct transitions. In the Fermi's golden rule implementation only excitations from $\epsilon_0$ to $\epsilon_0 + \Omega$ are possible. In contrast, in the perturbation theory implementation, transitions are possible into a broader energy range because of the finite time of the laser pulse. 
Generally speaking the energy integrated transition probability in Eq.~\eqref{eq:p01} grows monotonically with time, making the energy-integrated  transition rate (Eq.~\ref{eq:lasertens}) always positive. However if we inspect the energy resolved quantities, we notice that, right after the laser pulse is switched on, the transition probability $p_{01}$ given by Eq.~\eqref{eq:p01} is very broad in energy as the laser field restricted to short times $[0,t]$ contains many frequency components. As the time passes the central peak height increases while the distribution gets narrower and the transition probability resembles more and more a Dirac-$\delta$ function that we would expect from Fermi's golden rule (see Fig. \ref{fig:laser_exc} lower panel). \footnote{We observe, that the central peak grows during the whole duration of the laser pulse while the narrowing happens only for times where the time-envelope of the pulse has a negative slope (i.e. for times $t>6$).}

This narrowing effect implies that the transition rate (Eq.~\eqref{eq:lasertens}) after being initially positive, becomes negative at later times on the tail of the peak, e.g.\ for $\epsilon=3$ in Fig.~\ref{fig:laser_exc} (lower panel). Therefore, after an initial excitation, the electrons that were excited at those energies will be returned to their original state in the lower band. For low frequency excitations, most of the allowed transitions will lie well within the upper Hubbard band. Therefore the excitation and de-excitation happening on the energy tails of the transitions will be heavily shadowed by the always positive transition rate at the central peak. 
However, for large laser frequencies, a larger fraction of the transitions will be happening only at the borders of the Hubbard bands. In this scenario transitions happening at the peak of the energy resolved transition rate will not be activated since they will fall outside of the density of states, and the excitation and de-excitatiton of electrons at the tail will be more evident.

Finally we emphasize that the choice of the laser implementation makes little difference on the overall thermalization dynamics (i.e. the relaxation times vary by less than 5\%). 
 
\subsection{Population dynamics} \label{chap:popDyn}

\mw{In Fig. \ref{fig:time_distr} we further compare the energy resolved electron 
population $F(\epsilon) A(\epsilon)$} 
at different times for Boltzmann with Fermi's golden rule laser transitions, Boltzmann with perturbation theory laser transitions, and non-equilibrium DMFT. The three methods give very similar electron-distributions and 
provide evidence for 
impact ionization, since high energy doublons (spectral weight at high energies in the upper Hubbard band) disappear while low energy doublons increase more strongly. However there are some differences that we will address in the following. 
First, we can see that the laser excitation around $\epsilon \approx 2.5$ for $U=3.0$ or around $\epsilon \approx 3$ for $U=3.5$  displays sharper features for Boltzmann with Fermi's golden rule laser transition than for the two other cases. This is because Fermi's golden rule assumes a sharp transition at the laser frequency $\Omega$ which is only well justified when the period of the laser frequency is short compared to the time-envelope of the pulse. Therefore we can see a smoothened laser excitation due to an energetic broadening of the laser-pulse for Boltzmann with laser transitions in first order perturbation theory, as well as for non-equilibrium DMFT.

Furthermore, one observes that the non-equilibrium DMFT distribution has a finite electron density within the gap. A part of this effect arises because the Mott-gap gets filled with electrons as the energy (or temperature) of the system increases. This is physics beyond the Boltzmann description, as it corresponds to a redistribution of spectral weight in the density of states which contradicts the assumption made in Eq.~\eqref{eq:dosDef}. However, it is important to note that is not the only reason for the presence of spectral weight within the bandgap in the computed non-equilibrium DMFT distribution. It is noteworthy that a finite electron density within the gap (see Ref.~\onlinecite{Werner2014}, not shown here) is reported even at $t=0$.
The latter is due to the fact that the calculation of the time-resolved photo-emission spectrum was performed by integration only over short time intervals: this results in a purely numerical broadening in the frequency space. Therefore, for a meaningful comparison we convolute the Boltzmann electron distributions with a Gaussian, where the width ($\sigma$) is chosen such that the electron density inside the gap is approximately the same as at $t=0$ in the non-equilibrium DMFT case calculation (not shown here; $\sigma=0.14$). For both cases ($U=3.0$ and $U=3.5$) the broadened electron distribution is much closer to the DMFT result (see Fig.~\ref{fig:distr_smooth}). 

\subsection{Scattering strength} \label{chap:scatStr}

So far we have not discussed the role of the scattering pre-factor $\alpha$ that gives the strength of the interaction between excitations and has been used as a free parameter. 
If $\alpha$ were derived by taking the bare Hubbard interaction $U$ for the vertex  $w(\bold k_0,\bold k_1,\bold k_2,\bold k_3)$ in a QPBE, or for the vertex $\tilde U$ in \eqref{msbkskcbswx}, we would get $\alpha_{\text{bare}} = \frac{2 \pi}{\hbar} U^2$ \mw{(see Appendix \ref{chap:derAlpha})}.
The values we obtain for $\alpha$ in order to fit the time-scales differ significantly from this simple 
value.
Not only the values are different from $\alpha_{\text{bare}} $, also the dependency on $U$ is reversed: $\alpha$  decreases with increasing $U$. The reason for this 
is
that we have used a strongly renormalized spectral density which means that 
part of the interaction is already included in the density of states. 
The 
parameter
$\alpha$ would then just be the residual interaction between excitations within this renormalized density of states. 
Another effect which is disregarded in the Boltzmann approach approach is the filling of the Mott-Hubbard gap due to the laser excitation. This effect is larger for intermediate $U$ (i.e. small band-gaps) and gives rise to additional scattering channels via the in-gap states, leading to a larger $\alpha$ when trying to represent the dynamics in a Boltzmann approach. Indeed, full DMFT simulations yield the most rapid thermalization for intermediate values of $U$.\cite{Eckstein2009} While this might explain the counterintuitive decrease of $\alpha$,  
how to obtain the correct pre-factor for \mb{QBE 
}  for strongly interacting systems is left for further investigation.

\subsection{Three-step thermalization} \label{chap:threeStep}
An advantage of the quantum Boltzmann equation 
compared to non-equilibrium DMFT is the possibility to simulate the whole thermalization process, not only a short time interval after the laser pulse. By studying the 
distribution function
at different times during thermalization, one finds a third, intermediate characteristic timescale.

 Right after the laser pulse, the system shows a strong deviation from its original Fermi-Dirac distribution (see Fig.~\ref{fig:time_distr}). In a first step, as already discussed above, the highly excited electrons produce impact ionization until there are no electrons with sufficient energy any more. This happens in the case $U=3.5$ and $\Omega=3 \frac{\pi}{2}$ over the characteristic timescale $\gamma=53$ (see TABLE \ref{tab:times}).

In a second step, the thermalization proceeds through scatterings that leave the doublon and holon numbers unchanged. 
One doublon (holon) can scatter with another doublon (holon) redistributing the energy within the upper (lower) Hubbard band, or one doublon can scatter with a holon which corresponds to an exchange of energy between the upper and lower Hubbard bands. These doublon- and holon-conserving scatterings are not affected by the gap, hence they take place on a much faster timescale than the long-time thermalization. Since the number of doublons remains unchanged, this additional time scale is not visible in the doublon dynamics shown in Fig.~\ref{fig:nTU3d0}.
  
Only on a much longer time scale ($\tau=254$ in the case $U=3.5$ and $\Omega=3 \frac{\pi}{2}$, see Tab.~\ref{tab:times}) the system reaches a full thermalization. This requires the creation of high energy doublons (holons) through multiple scattering processes so that impact ionization can  eventually thermalize the number of doublons (holons). These processes constitute the second, long time scale in Fig.~\ref{fig:nTU3d0}.

 We now systematically analyze this intermediate thermalization step by fitting two Fermi-Dirac functions within  the lower and upper Hubbard band, respectively, to the distribution function \mw{$F(t,\epsilon)$} in every time step.
 Such independently thermalized distributions in the upper and lower band are common in semiconductor physics, and have also been observed in more strongly correlated insulators using non-equilibrium DMFT for the ionic Hubbard model.\cite{Dasari2018} The fit yields two  chemical potentials $\mu$,  two inverse temperatures $\beta$
and two (squared)  deviations $\Delta$ from the Fermi-Dirac functions for the lower and upper bands \mw{
\begin{subequations}
\begin{align}
\Delta_\textrm{lower}(t) & \equiv \int_{-\infty}^{-\frac{E_\textrm{gap}}{2}} \!\!\!  \mathrm d \epsilon {\Big( F(t,\epsilon) -  \frac{1}{e^{(\epsilon-\mu_\textrm{lower}(t) ) \beta_\textrm{lower}(t)}  + 1} \Big)}^2 \!\! \textrm{,}\\
\Delta_\textrm{upper}(t) & \equiv \int_{\frac{E_\textrm{gap}}{2}}^{\infty} \!\! \mathrm d \epsilon {\Big( F(t,\epsilon) -  \frac{1}{e^{(\epsilon-\mu_\textrm{upper}(t) ) \beta_\textrm{upper}(t)}  + 1} \Big)}^2 \textrm{.}
\end{align}\label{eq:fittingerror}
\end{subequations}}

\begin{figure}
\includegraphics[width=7.5cm]{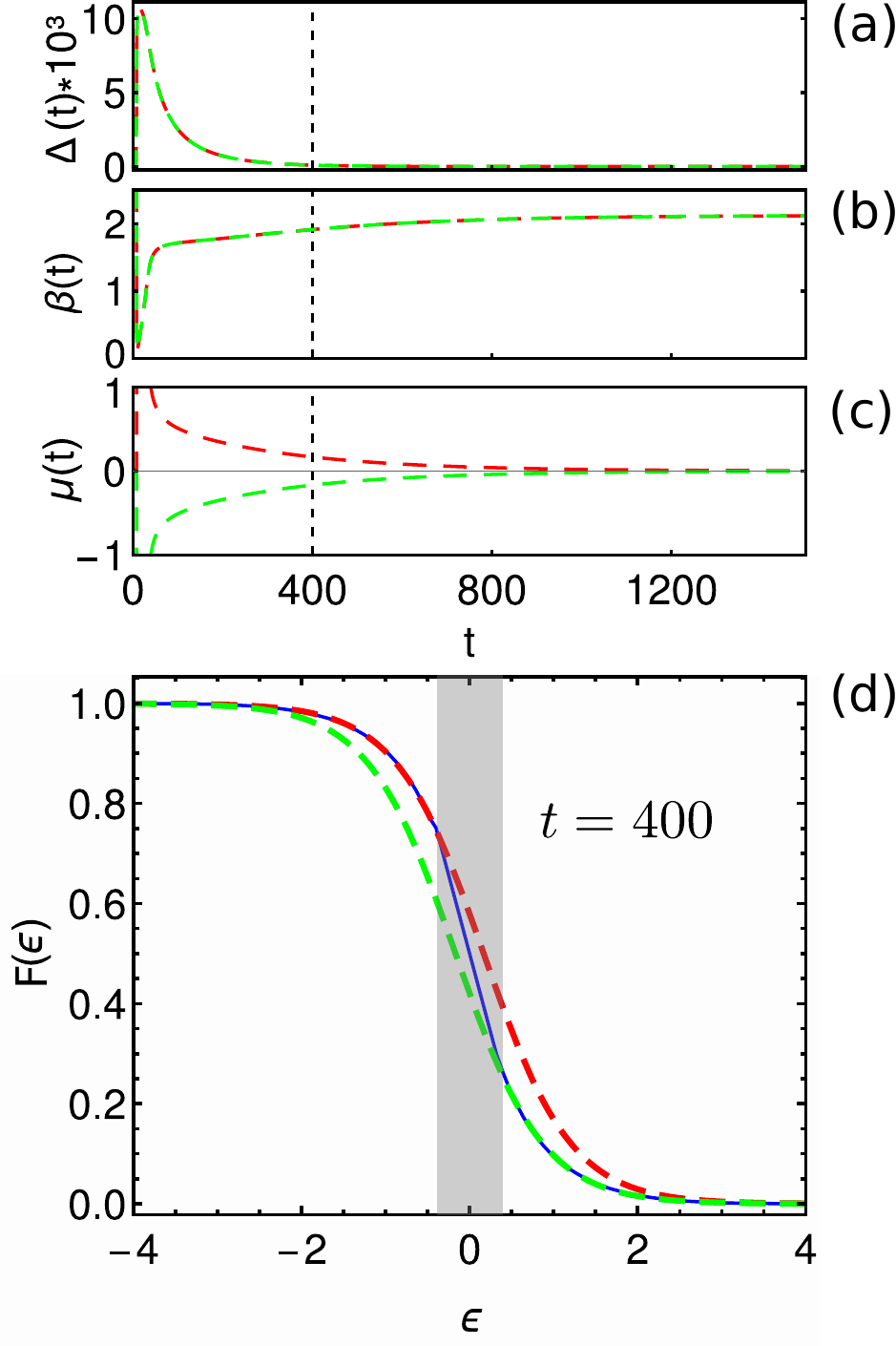}
\caption{Fermi-Dirac fit to the non-equilibrium distribution function within the lower (red) and the upper (green) Hubbard-band. 
The interaction parameter for this case is $U=3.5$ and the laser frequency is $\Omega=3 \frac{\pi}{2}$. The fitting error (a) [Eq.~\eqref{eq:fittingerror}] has almost vanished at time $t=400$ (black, dashed line) whereas the inverse temperatures (b) are not yet thermalized and  the chemical potentials (c) still differ significantly.  Panel (d) shows the actual distribution function (blue, solid line) in comparison to the Fermi-Dirac fit in the lower (\mw{red}) and upper (\mw{green}) Hubbard band for $t=400$. The gray area marks the gap of size $E_g=0.8$.
 }\label{fig:thermalization}
\end{figure}

 Figure~\ref{fig:thermalization} (a) shows that the deviation from a Fermi-Dirac distribution is largest directly after the laser pulse. It decays over a characteristic timescale of $\eta=100$ and is essentially zero at $t=400$. Because the system is particle-hole symmetric, the inverse temperatures in
 Fig.~\ref{fig:thermalization} (b) are equal at all times for both Hubbard bands but at $t=400$ still lower than the equilibrium (long time) value. Even more obvious is the substantial difference in the two chemical potentials at this intermediate time in  
 Fig.~\ref{fig:thermalization} (c). Let us stress once more that one cannot see the intermediate relaxation stage, where the upper and lower Hubbard bands are thermalizing independently, in the time dependence of the double occupation.


The third thermalization step, full thermalization between the bands (involving further doublon-holon generation), is clearly seen in Fig. \ref{fig:thermalization}(c) as the equalization of the two chemical potentials.

%

\section{Conclusion and outlook} \label{chap:concl}

We have \mb{provided a NEGF and DMFT based derivation of how to construct a quantum  Boltzmann equation for the electron population dynamics in strongly correlated systems. 
The equation is structurally identical to the quasiparticle Boltzmann equation with dropped momentum conservation, but more generally applicable even if well defined quasiparticles do not exist.  The basic assumption is that the equilibrium DMFT spectral function $A(\omega)$ does not change due to the non-equilibrium dynamics but only the distribution function $F(t,\omega)$.
}

 We applied the quantum Boltzmann equation  
 to a system in its Mott-insulating phase, where well defined quasiparticles do not exist, and 
obtained a good qualitative and semi-quantitative agreement with non equilibrium DMFT results.

This was shown by a direct comparison to the thermalization dynamics obtained from non-equilibrium DMFT. Both the  features of impact ionization, a characteristic feature of early thermalization dynamics, and the  ratio between the long-time thermalization time and the impact ionization time are very close to the non-equilibrium DMFT values. Not only the timescales coincide, but the electron distribution itself is in good agreement with the non-equilibrium DMFT distributions at different times. We find that even for a Mott insulator the dominant factor for the thermalization is the phase- (energy-) space that is available for scattering as this is exactly what is described by the Boltzmann 
term. The applicability of the \mw{quantum Boltzmann}\old{approach} 
\new{equation} is however limited to cases where the spectral function is not significantly modified by the excitation, as for instance in the case of temperature-induced filling of the Mott-Hubbard gap or a dynamically induced insulator-metal transition. 

Additionally we have implemented two different laser-excitation models, with and without the inclusion of coherent excitation processes. We found that the inclusion of coherent effects in the implementation of the laser excitation term captures important details of the DMFT simulations such as oscillations with two times the laser frequency or a transient increase in the double occupancy. We have also been able to pinpoint the physical mechanism behind the oscillations and the overshooting of the doublon population. 

Finally we were able to observe an intermediate phase where the upper and lower Hubbard bands are populated according to Fermi-Dirac distributions with the same temperature but different chemical potentials. 
Such an analysis 
could not be performed on the 
non-equilibrium DMFT data 
since 
the accessible timescales were too short. 

We conclude that, with a few caveats, an appropriately constructed Boltzmann formalism is able to qualitatively and semi-quantitatively describe thermalization dynamics even in strongly correlated materials. This is useful as the computational cost of Boltzmann simulations is much lower than that of non-equilibrium DMFT making it possible to simulate longer timescales, and 
to study the 
thermalization process. Further, its simplicity may help to understand the physical processes observed.

\new{
As an {outlook} we would like to discuss two aspects. First, the QBE is applicable in many additional situations. For a weakly correlated metal, the conventional Boltzmann equation with $\rho(\omega)$ instead of the interacting $A(\omega)$ is recovered, and the scattering amplitude is given by the perturbative result discussed in Appendix \ref{chap:derAlpha}. If we add essentially only a single particle at energy $\omega$, phase-space arguments imply a Fermi liquid relaxation rate $\sim \omega^2$ and $\sim T^2$. More interesting is the strongly correlated metallic phase with a narrow quasiparticle peak. Here, we expect reasonably good results as long as our assumption that $A(\omega)$ is not altered by the non-equilibrium dynamics holds.
}

\new{
Second, it is intriguing to extend the formalism also to the case of a finite electric field, calculating transport properties. Here, we  additionally need the Fermi velocity ${\mathbf v}({\mathbf k})$ in the Boltzmann equation (\ref{eq:boltzmann}). In the spirit of (optical) conductivity calculations \cite{Assmann2016} in density functional theory (DFT)+DMFT \cite{Anisimov1997,Lichtenstein1998,Held2006,Kotliar2006,Held2007} we might assume here that the dipole matrix element is still given by the non-interacting  wave function or the Peierls approximation $\bold v (\bold k)   = \frac{1}{\hbar} \nabla_{\bold k} \epsilon(\bold k)$.
But since the electric field is pointing in a certain direction and considering that $\nabla_{\bold k} \epsilon(\bold k)$ is often strongly 
anisotropic, 
approximating  $F(\bold k , t, \omega)$ by $F(t, \omega)$ appears to be a too crude approximation. A possibility is to consider the  momentum ${\mathbf k}_\parallel$ parallel and ${\mathbf k}_\perp$ perpendicular to the electric field and using only for the latter an energy- instead of  ${\mathbf k}_\perp$-dependent distribution function. Of course such a transport calculation also neglects vertex corrections unless these are included already in the scattering amplitude $\alpha$.}

\paragraph*{Acknowledgments}
We thank Anna Kauch, Enrico Arrigoni, Hans Gerd Evertz, Marcus Kollar, Florian Maislinger, and Max Sorantin, for valuable discussions.
 This work has been supported in part by the  European Research Council 
under the European Union's Seventh
Framework Program (FP/2007-2013) through  ERC grant agreement n.\ 306447, 
ERC grant agreement n.\ 724103, 
ERC starting grant No. 716648,
the Austrian Science Fund (FWF) through Doctoral School W1243 Solids4Fun (Building Solids for Function) and SFB ViCoM F41. M. Battiato acknowledges the Austrian Science Fund (FWF) through Lise Meitner position M1925-N28 and Nanyang Technological University, NAP-SUG for funding.

\appendix
\section{Basis functions} \label{chap:appBasis}
We use piecewise polynomial,  discontinuous basis functions. For defining these basis functions and working with them we will not use a global index running over all basis functions (as in section \ref{chap:numImp}) but two indices $\{I,i\}$ and we write the corresponding basis function as $\Phi_I^i(x)$.
\mb{
We subdivide the energy range} into $N_{E}$ mesh elements with lower boundaries $a_I$ and upper boundaries of $b_I$ with $I \in [1,N_E]$ and $a_I = b_{I-1}$. Each basis function is zero everywhere except within its corresponding element. Inside the element the basis function is equal to a normalized Legendre polynomial of order $i$. We can write this  as  
\begin{equation}
 \Phi_{I}^{i}(x)= \left\{
\begin{array}{ll}
    \sqrt{\frac{2 i +1}{b_I - a_I}} P_i ( \frac{2 x - b_I - a_I}{b_I - a_I} ) & \quad  a_I \leq x < b_I \\
    0 & \quad \textrm{otherwise}
\end{array} \right. ,
\end{equation}
where $P_i(y)$ is the Legendre polynomial of order $i$ at position $y$.

This basis has two main advantages. First, it allows for discontinuities which makes it easier to cover the steep changes in population typical of the Fermi-Dirac at low temperature. The second advantage is that these basis functions have a small support while still covering the full space and being orthogonal. In combination with the $\delta$-function that ensures energy conservation in the scattering tensor [Eq.~\eqref{eq:scat}] this ensures the maximum sparsity of the scattering tensor. We use basis functions up to second order (i.e. $i \in [0,2]$) and $N_E = 16$ ($N_E = 14$) mesh elements for $U=3$ ($U=3.5$) simulations.  

\section{Derivation of $\alpha$ for small $U$ in the Hubbard model} \label{chap:derAlpha}
The scattering strength $\alpha$ in Eq.~\eqref{eq:col3} is a priori unknown for a given interaction $U$. If the interaction is weak (i.e. the spectral density is approximately the non-interacting density of states, $A(\epsilon) \approx \rho(\epsilon)$), we are able to calculate it using Fermi's golden rule following the standard procedure. In the paper we treat a Hubbard model on a hypercubic lattice for which the Hamiltonian reads
\begin{equation}
\hat {\mathcal H} = \underbrace{-t \sum_{\langle i,j \rangle, \sigma} \hat {a}^\dagger _ {i \sigma} \hat{a}_{j \sigma}}_{\hat {\mathcal H_0}} + \underbrace{U \sum_i \hat {a}^\dagger _ {i \uparrow} \hat{a}_{i \uparrow} \hat {a}^\dagger _ {i \downarrow} \hat{a}_{i \downarrow}}_{\hat {\mathcal V}} \label{eq:hubbard}
\end{equation}
with the hopping amplitude $t$ and the on-site interaction $U$. The operator $\hat {a}^\dagger _ {i \sigma}$ ($\hat{a}_{i \sigma}$) creates (annihilates) an electron with spin $\sigma$ at site $i$ and $\sum_{\langle i,j \rangle}$ means that the index $i$ runs over all sites while $j$ only counts the sites neighboring to $i$. 

The first term in Eq.~\eqref{eq:hubbard} is diagonalized by the introduction of new creation (annihilation) operators $c^\dagger_{\bold k \sigma}$ ($c_{\bold k \sigma}$) with
\begin{equation} \label{eq:Kdef}
\hat c_{\bold k \sigma} \equiv  \frac{1}{\sqrt{N}} \sum _ j e^{ i \bold R_j \cdot \bold k}  \hat{a}_{j \sigma} \textrm{,} \quad \hat c^\dagger _{\bold k \sigma} \equiv  \frac{1}{\sqrt{N}} \sum _ j e^{- i \bold R_j \cdot \bold k}  \hat{a}^\dagger_{j \sigma}
\end{equation}
that create (annihilate) an electron with wave vector $\bold k$ and spin $\sigma$. The quantity $N$ denotes the number of lattice sites. As the $\hat c$-operators diagonalize the non-interacting part of the Hamiltonian ($\hat{\mathcal H_0}$) they describe well-defined quasiparticles in the case $U \ll t $. We will use this knowledge in the next step to calculate the correct pre-factor of the collision term.

Let us first write the collision operator for the system described above,
\begin{equation}
\begin{split}
\frac{\partial f_{\sigma_0}(\bold k_0)}{\partial t} =  \frac{1}{2}& \sum_{\underset{\sigma_1, \sigma_2, \sigma_3}{\bold k_1, \bold k_2, \bold k_3}}  \Big [ \mathcal W _{\sigma_0 \sigma_1 \sigma_2 \sigma_3 }(\bold k_0,\bold k_1,\bold k_2,\bold k_3) \\
& \times \mathcal{P}[f_{\sigma_0}(\bold k_0), f_{\sigma_1}(\bold k_1),f_{\sigma_2}(\bold k_2),f_{\sigma_3}(\bold k_3)]  \Big ],
\end{split} \label{eq:collAp}
\end{equation}
where the factor $\frac{1}{2}$ in front of the sum is needed to prevent double counting. The factor $\mathcal W _{\sigma_0 \sigma_1 \sigma_2 \sigma_3 }(\bold k_0,\bold k_1,\bold k_2,\bold k_3)$ is the transition rate from a state with two electrons  $\ket{i}  \equiv \hat c^\dagger _{\bold k_0 \sigma_0} \hat c^\dagger _{\bold k_1 \sigma_1} \ket{0}$ to a state $\ket{f}  = \hat c^\dagger _{\bold k_2 \sigma_2} \hat c^\dagger _{\bold k_3 \sigma_3} \ket{0}$ (or the inverse process). With Fermi's golden rule one can calculate this rate as 
\begin{equation} \label{eq:transR1}
\mathcal W _{\sigma_0 \sigma_1 \sigma_2 \sigma_3 }(\bold k_0,\bold k_1,\bold k_2,\bold k_3) = \frac{2 \pi}{\hbar} T(t) |  \bra{f} \hat {\mathcal V} \ket{i} |^2
\end{equation}
with the function
\begin{equation}
T(t) = \frac{ \sin \left ( (\epsilon_0 + \epsilon_1 - \epsilon_2  - \epsilon_3) t \right ) }{ \pi \left (\epsilon_0 + \epsilon_1 - \epsilon_2  - \epsilon_3 \right )} \textrm{ .}
\end{equation}
Using the anti-commutation relations for fermionic creation and annihilation operators we can evaluate Eq.~\eqref{eq:transR1} and get (with the indices and dependencies of $\mathcal W$ dropped)
\begin{equation} \label{eq:transRateApp}
\mathcal W = \frac{2 \pi}{\hbar} T(t) \frac{U^2}{N^2}  \delta_{\sigma_0  \bar \sigma_1}  \delta_{\sigma_2  \bar \sigma_3}
  \sum_{\bold G} \delta _{\left ( \bold k_{0} + \bold k_1 - \bold k_2 - \bold k_3  \right ), \bold G}
\end{equation}
where $\bar \sigma \equiv -\sigma$ and $\sum_{\bold G}$ means the sum over all reciprocal lattice vectors $\bold G$. 
The two spin-deltas $\delta_{\sigma_0  \bar \sigma_1}  \delta_{\sigma_2  \bar \sigma_3}$ represent the fact that spin is conserved in the scattering event and that only electrons with opposite spin may scatter with each other, which follows from the purely local nature of the interaction. As the system is spin degenerate we do not need to distinguish between the distributions for spin up and down, i.e. $ f(\bold k_i) \equiv  f_{\uparrow}(\bold k_i)=  f_{\downarrow}(\bold k_i)$. 
If the volume of the system is large  we are allowed to use integrals instead of sums over $\bold k$ in Eq.~\eqref{eq:collAp} ($\sum_{\bold k_i} \to \frac{N}{V_{BZ} } \int_{BZ} \mathrm d ^d k_i$). The Kronecker-delta then becomes a delta function ($\delta _{\left ( \bold k_{0} + \bold k_1 - \bold k_2 - \bold k_3  \right ), \bold G}  \to \frac{V_{BZ}}{N} \delta \left (\bold k_{0} + \bold k_1 - \bold k_2 - \bold k_3   - \bold G \right ) \equiv \frac{V_{BZ}}{N} \delta _{\bold k}$) and the $T(t)$-function may be replaced by a delta function in the case of sufficiently large times ($T(t) \to \delta \left ( \epsilon(\bold k_0) + \epsilon(\bold k_1) - \epsilon(\bold k_2) - \epsilon(\bold k_3)  \right ) \equiv \delta _{\epsilon}$). Eq.~\eqref{eq:collAp} then becomes 
\begin{equation}
\begin{split}
&\frac{\partial f (\bold k_0 )}{\partial t } = \sum_{\bold G} \int \mathrm d ^d k_1 \mathrm d ^d k_2 \mathrm d ^d k_3  \Big [ \frac{2 \pi}{\hbar} \frac{U^2}{{V_{BZ}}^2} \delta_{\bold k} \delta_{\epsilon} \\
 &\times \mathcal{P}[f(\bold k_0), f(\bold k_1),f(\bold k_2),f(\bold k_3)] \Big ]
\end{split} \label{eq:collAp3}
\end{equation}
and from comparison with Eqs.~\eqref{eq:col} and \eqref{eq:scatAmpl} the weak-interaction limit for the scattering probability follows as $w(\bold k_0,\bold k_1,\bold k_2,\bold k_3) = \frac{2 \pi}{\hbar} \frac{U^2}{{V_{BZ}}^2}$. Therefore $\alpha$ becomes
\begin{equation}
\alpha = \frac{2 \pi}{\hbar} U^2 \equiv \alpha_\textrm{bare} \textrm{ .}
\end{equation}


%
\end{document}